\documentclass[fleqn,usenatbib]{mnras}

\usepackage{newtxtext,newtxmath}
\usepackage{graphicx}
\usepackage{hyperref}
\usepackage[defaultcolor=red]{changes}

\usepackage[T1]{fontenc}

\DeclareRobustCommand{\VAN}[3]{#2}
\let\VANthebibliography\thebibliography
\def\thebibliography{\DeclareRobustCommand{\VAN}[3]{##3}\VANthebibliography}
\newcommand{\EA}{E+A}

\usepackage{graphicx}	
\usepackage{amsmath}	
\usepackage{cprotect}

\newcommand{\ha}{\textrm{H}\ensuremath{\alpha}}
\newcommand{\hb}{\textrm{H}\ensuremath{\beta}}

\newcommand{\oiii}{[\textrm{O}~\textsc{iii}]}
\newcommand{\nii}{[\textrm{N}~\textsc{ii}]}

\graphicspath{{./}{figures/}}

\title[PAE-based anomaly detection for galaxy spectra]{Fast and efficient identification of anomalous galaxy spectra with neural density estimation}

\author[]{
Vanessa B\"ohm$^{1,2}$\thanks{E-mail: vboehm@berkeley.edy},  
Alex Kim$^{2}$ and St\'ephanie Juneau$^{3}$
\\
$^{1}$Berkeley Center for Cosmological Physics, University of California, Berkeley, CA 94720, USA\\
$^{2}$Lawrence Berkeley National Lab, 1 Cyclotron Road, Berkeley, CA 94720, USA\\
$^{3}$NSF's NOIRLab, 950 N Cherry Avenue, Tucson, AZ 85719, USA
}

\date{Accepted XXX. Received YYY; in original form ZZZ}

\pubyear{2023}

\begin{document}
\label{firstpage}
\pagerange{\pageref{firstpage}--\pageref{lastpage}}
\maketitle

\begin{abstract}
Current large-scale astrophysical experiments produce unprecedented amounts of rich and diverse data. This creates a growing need for fast and flexible automated data inspection methods. Deep learning algorithms can capture and pick up subtle variations in rich data sets and are fast to apply once trained. Here, we study the applicability of an unsupervised and probabilistic deep learning framework, the Probabilistic Autoencoder (PAE), to the detection of peculiar objects in galaxy spectra from the SDSS survey. Different to supervised algorithms, this algorithm is not trained to detect a specific feature or type of anomaly, instead it learns the complex and diverse distribution of galaxy spectra from training data and identifies outliers with respect to the learned distribution. We find that the algorithm assigns consistently lower probabilities (higher anomaly score) to spectra that exhibit unusual features. For example, the majority of outliers among quiescent galaxies are E+A galaxies,
whose spectra combine features from old and young stellar population. Other identified outliers include LINERs, supernovae and overlapping objects. Conditional modeling further allows us to incorporate additional information. Namely, we evaluate the probability of an object being anomalous given a certain spectral class, but other information such as metrics of data quality or estimated redshift could be incorporated as well. We make our \href{https://github.com/VMBoehm/Spectra_PAE}{code} publicly available.
\end{abstract}

\begin{figure*}
\includegraphics[scale=0.45]{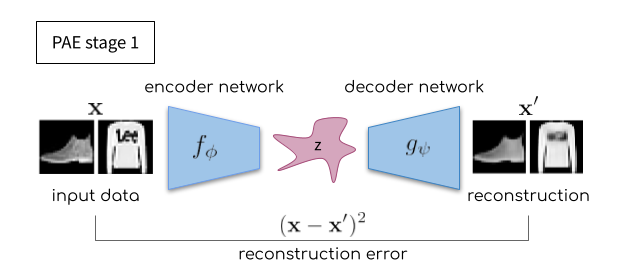}
\includegraphics[scale=0.45]{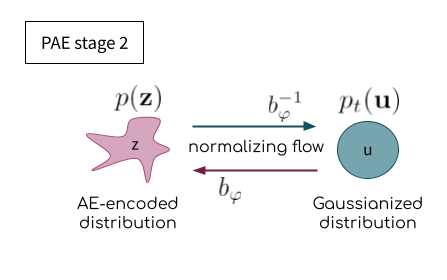}
\caption{\label{fig:1} The two-step training of a Probabilistic Autoencoder (PAE) illustrated on an image data set. In the first stage, an autoencoder (AE) consisting of two neural networks is trained to minimize the reconstruction error after compressing (encoding) the data in a lower-dimensional latent space and decompressing (decoding) it back into the high-dimensional data space. In the second stage, a normalizing flow (NF) is trained to learn a bijective mapping from the AE-encoded latent space to a space in which the encoded data follows a Gaussian distribution.}
\end{figure*}
\section{Introduction} 
\label{sec:intro}
Current and upcoming stage-4 astronomical surveys are about to provide astronomers with an overwhelming amount of data. Among them are the Dark Energy Spectroscopic Instrument (DESI) Survey~\citep{DESI}, which is in the process of taking high-resolution spectra of tens of millions of galaxies. 
In light of the exponential increase in the size of astronomical data sets, there is a need for methods that allow for fast and efficient online inspection of data samples.

In particular, methods that can identify and flag candidates for follow-up observations. The most prominent examples are transients such as supernovae, kilonovae, or tidal disruption events. 
Machine learning algorithms are extremely fast to evaluate once an upfront computational cost in training has been paid. They are further known for their generalization properties and ability to extract information from complex and high-dimensional data. Because of these properties, they are a natural candidate for the task. In the paper, we explore the use of unsupervised machine learning to mine spectral data sets for anomalous objects.

Deep learning approaches can be broadly grouped into two categories- supervised and unsupervised (semi- and self-supervised approaches exist, too). In supervised learning we have access to a training set for which the solution of the task to be learned is known and we train the algorithm to reproduce the known results. In astronomy, the labeled training data set is often produced by means of data simulations. 
The generalization properties of neural networks guarantee that the algorithm can be applied to new unseen data. Supervised learning has the advantage that it provides a label for the identified object and that it is extremely accurate at identifying objects similar to ones it has seen during training. However, supervised learning can miss objects that are atypical or simply have slightly different properties to the ones it was trained on. 

A complementary approach is to use unsupervised learning to identify special objects. Here, the task becomes a type of anomaly or out-of-distribution detection. This approach is extremely general and less targeted, but can capture a wide range of anomalies, including unanticipated ones. These approaches do not require a labeled training set and can be trained on the actual data. The advantage of training on the real data is that the algorithm will not be sensitive to potential differences between the real data and data simulations. In this paper we explore the second route and suggest ways in which the identified anomalies can be organized and analyzed.  

Reliable and efficient anomaly detection in very large and rich data sets is an extremely active field of research in many domains, including astronomy~\citep{2017MNRAS.465.4530B, 2021ApJS..255...24V, Stein_2022}, high energy physics~\citep{PhysRevD.101.075021, Blance2019,Cerri2019}, and computer science~\citep{AnomalyReview1, AnomalyReview2}. Proposed methods cover a wide range of machine learning models and anomaly metrics ranging from the reconstruction error of autoencoders~\citep{PhysRevD.101.075021}, over distance estimates obtained from Random Forests~\citep{2017MNRAS.465.4530B} to density estimates obtained from Variational Autoencoders or Normalizing Flows~\citep{Nalisnick2019, Nalisnick2019b}. While all of these methods have been shown to work in some contexts, each of them is also subject to certain limitations. For example, the anomaly detection accuracy of the reconstruction error of AEs depends on the latent space dimensionality and expressiveness of the employed neural network. A powerful decoder network combined with a latent space of high enough dimensionality is able to reconstruct even out-of-distribution samples with small reconstruction error. Random Forest estimates have achieved impressive results in the context of finding `weird' galaxy spectra and identifying transients, but their results are sometimes hard to interpret and it is unclear whether they cover the entire space of anomalies. 

In this paper, we briefly investigate the reconstruction error method but we focus on the third approach -- density estimation. 
In the context of density estimation, anomalies are supposed to have low probability under a probability density estimate obtained from training data. Until recently, flexible density estimation was mostly limited to the method of Kernel Density Estimation (KDE), which is computationally intractable in high dimensions and for large data samples and has limited flexibility due to the restricted space of kernel functions. However, recent advancements in neural network-based density estimation have produced models such as normalizing flows~\citep{NF_review} or autoregressive flows~\citep{MAF}, which are able to efficiently learn probability densities of relatively high-dimensional and complex data. This is demonstrated through their ability to sample extremely realistic data realizations from these distributions~\citep{glow}.

Despite their successes in data generation, neural density estimators in their vanilla form have been found to be inadequate for or even exhibit catastrophic failure in anomaly detection~\citep{Nalisnick2019}. A number of adapted models, however, have been able to resolve these issues and have demonstrated excellent separations between in- and out-of-distribution data on standard test examples~\citep{LikelihoodRatioAI, Nalisnick2019b}. Here, we use a Probabilistic Autoencoder (PAE)~\citep{PAE}, which removes troublesome singular dimensions through data compression before applying a neural density estimator in the compressed space. This approach is well suited for the galaxy spectra, which are well known to reside on a lower dimensional manifold, allowing for almost lossless compression to comparatively low dimensionalities~\citep{2012MNRAS.421..314C}.

The PAE framework has been successfully applied to spectroscopic astrophysical data by~\citet{Stein_2022} to learn the intrinsic diversity of Type Ia supernovae from a sparse spectral time series and more recently by~\citet{Pat_2022} to study intrinsic degeneracies in galaxy spectra. ~\citet{2020AJ....160...45P} studied the spectral components learned by a Variational Autoencoder (VAE). Compression through autoencoding (AE) is also part of the recently published SPENDER framework for analyzing, representing, and creating galaxy spectra~\citep{2022arXiv221107890M,Liang+2023}. 

This paper is organized as follows: We start by describing the data set and data preparation. This is followed by a general description of the anomaly detection method and how it is adapted to fit the specific spectroscopic data set. We take great care to explain our choices for the neural network architecture, test the robustness of the method to these choices and describe the exact training procedure for reproducibility. Results are presented in Section~\S\ref{sec:OOD} together with a comprehensive analysis of the identified outliers. We conclude in Section~\S\ref{sec:conclusions}.
\section{Data set and data preparation}
\label{sec:data}
We use publicly available data from the SDSS-BOSS DR16 release
\citep{2000AJ....120.1579Y,
2002AJ....124.1810S,
2002AJ....123.2945R,
2006AJ....131.2332G,
2020ApJS..249....3A}
and employ a number of data cuts in order to ensure a minimum data quality and ease comparison with prior work. In particular, our
target selection mirrors that
of \citet{2020AJ....160...45P}, who
selected from the main galaxy and quasar samples.
In addition we require the data come from
plates with {\tt PLATEQUALITY="good"}. The resulting set contains only objects that have been classified as galaxy (GAL) or quasar (QSO) by the SDSS pipeline.
\footnote{
The query we use is
{\tt SELECT plate, mjd, fiberid
FROM SpecObj 
WHERE 
 ((legacy\_target1 \& (2+4+64)) > 0) AND (z > 0) AND zwarning=0}, which is cut off
 by the SDSS SkyServer 500,000 row limit.
}

We preprocess the data by de-redshifting into the rest frame using redshift estimates obtained by the RedRock algorithm
\citep{2012AJ....144..144B}. We further rebin the spectra onto a logarithmically spaced grid with $D{=}1000$ bins ranging from $3388\textup{\AA}$ to $8318\textup{\AA}$. We note that these choices are somewhat arbitrary and that higher resolutions or different minimal and maximal wavelengths could be used. The new bins contain the average of previous pixel values, and bins containing no measurements are masked. The amplitude of each spectrum is re-scaled to a luminosity distance corresponding to a redshift of $z{=}0.1$ in the Planck 2018 best fit cosmology
\citep{2020A&A...641A...6P}.  

We restrict the data set to redshifts $0.05{<}z{<}0.36$, remove spectra with masked fractions $f_m{>}0.4$, and add a noise-floor to pixels with signal-to-noise $S/N{>}50$, such that the signal-to-noise of any pixel never exceeds 50. This last step helps ease the network training, since high S/N pixels are likely to dominate the training loss, while not necessarily being the most relevant for the task. 
We further only include spectra with a cumulative signal-to-noise $S/N{>}50$ in our data set.
This cut removes objects on
the extreme tail of the signal-to-noise distribution, which goes beyond what would be 
expected from the
magnitude limit of SDSS sample selection ($S/N{\gtrsim}200$ in our redshift range).
The purpose of removing low-quality data is to ensure that the algorithm focuses on outliers that are anomalous because of the properties of the observed spectrum and not because of anomalously low data quality. While the algorithm could also be used to identify glitches in the reduction pipeline or other problems such as a misaligned fiber, it makes sense to remove quality issues that can be identified through other means.

To facilitate the training of the neural networks  we further divide the data by their pixel-wise mean value before feeding it into the networks. The total size of the data set after cuts is $N{=}$349,104, which we split into a training ($N_\mathrm{train}{=}$209,462) and test set ($N_\mathrm{test}{=}$139,642), respectively.
The training set is used to train and calibrate the algorithm. Once the training is completed, we run the algorithm on the test sample. We present results obtained for the test sample in \S\ref{sec:OOD}. 

We additionally obtained spectral measurements for the test set from the Portsmouth Stellar Kinematics and Emission Line Fluxes value-added catalog \citep{Thomas+2013}. This catalog was last generated for SDSS DR12, and we found matches for 136,876 galaxies (98\% of 139,642) based on the plate, MJD (modified Julian date) and fiberid identifiers. For those cases, we retrieved \ha\ emission line flux and equivalent width as well as the stellar velocity dispersion ($\sigma_{\textrm{stars}}$) and their measurement uncertainties. On average, \ha\ equivalent width is a proxy for the specific star formation rate of galaxies \citep[star formation rate divided by the total stellar mass;][]{Brinchmann+2004} while $\sigma_{\textrm{stars}}$ is a proxy for the mass enclosed within the SDSS fiber aperture. These quantities will be used to assess general galaxy population properties and verify our selection of ``normal'' quiescent galaxies in \S\ref{sec:galpop} but not for any precise quantitative analyses. 

\section{Methods}
\label{sec:method}
Our anomaly detection algorithm is based on a deep machine model, the probabilistic autoencoder (PAE)~\citep{PAE}. We start by a description of this algorithm in its vanilla form and then discuss the adaptions made for the specific task at hand.

\subsection{Introduction to Probabilistic Autoencoder}
\begin{figure*}
\includegraphics[width=0.98\textwidth]{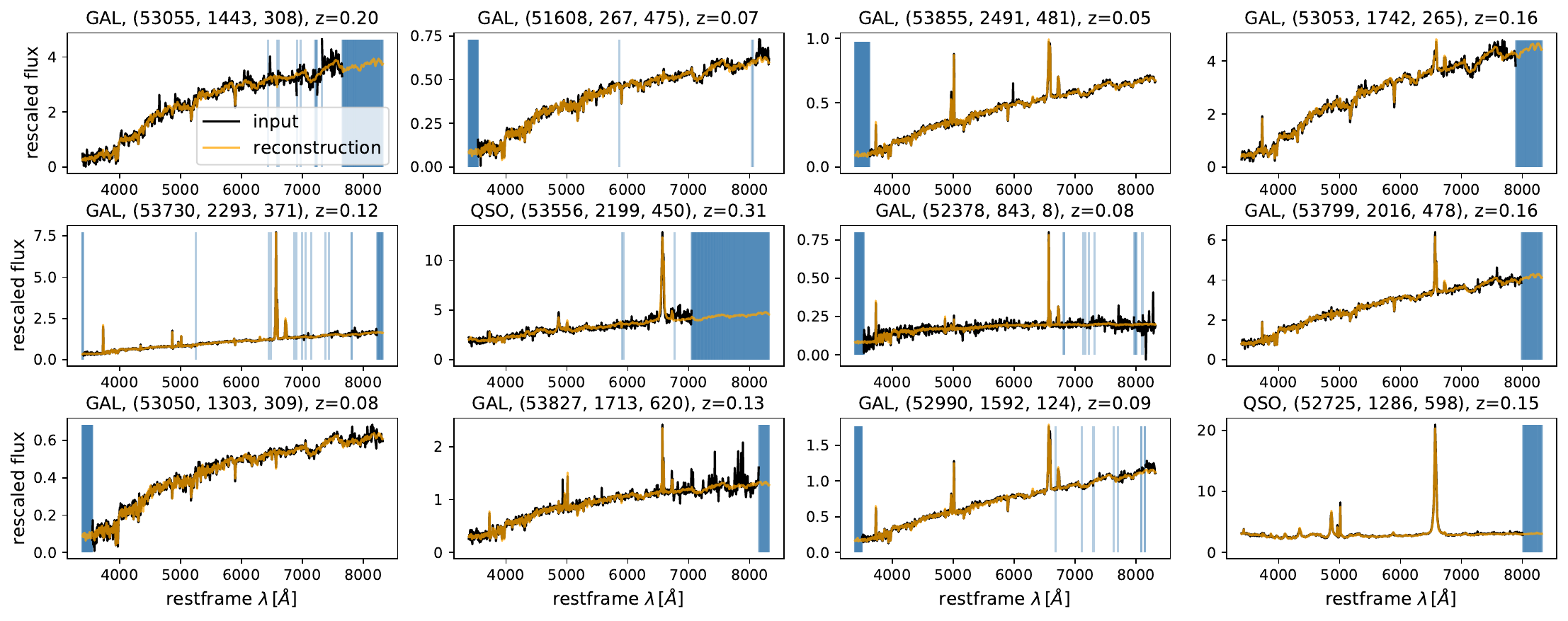}
\caption{\label{fig:2} Re-scaled noisy and incomplete galaxy spectra from the test data set (black) and their denoised and inpainted reconstructions (orange). The spectra have been randomly selected. Masked pixels are indicated by blue lines. We show the MJD number, plate and fiber id, as well as the estimated redshift of each spectrum in the panel titles.}
\end{figure*}
The probabilistic autoencoder is a two-stage probabilistic machine learning model composed of an autoencoder and a normalizing flow. An autoencoder is a dimensionality reduction algorithm, and a normalizing flow is a neural density estimator. The two stages, autoencoder (AE) and normalizing flow (NF) are trained separately. 
The AE consists of two approximately symmetrical neural networks. The encoder network, $f_\phi$, with trainable network parameters $\phi$, maps the high dimensional data into a lower dimensional latent space, $z$. The decoder network, $g_\psi$, with trainable network parameters $\psi$, decompresses latent space points and maps them back into data space. The two networks are trained jointly. The training objective is to minimize the average reconstruction error or mean squared error (MSE) between the input data $x$ and the decompressed data $x'$. The loss function that is minimized during training is,
\begin{equation}
    \mathcal{L}_{\mathrm{AE}}(\phi,\psi)=\underset{x \sim p(x)}{\mathbb{E}}\left[\frac{1}{D}\sum_{i=0}^{D-1}\left(x_i-g_{\psi}[f_\phi(x)]_i\right)^2\right],
\end{equation}
where the index $i$ is used to label individual pixels in the data. 
The dimensionality of the AE encoded space is $K{\ll}D$. 

Autoencoders are sometimes described as a non-linear generalization of principal component analysis (PCA). Different to a PCA, an autoencoder does not enforce a diagonal covariance in encoded space. Because AEs leverage non-linear neural network-based transformations, the AE compressed space can be irregular and its exact shape can depend on the random initial conditions of the training process. This dependence is cured by the density estimation in the second step of the PAE training. We verified that our results are indeed robust to changing random variables, such as the randomly drawn initial values of the network parameters and the order of data samples presented to the algorithm during training. We choose a Probabilistic Autoencoder over altenative models, such as the Variational Autoencoder~\citep{ KingmaWelling13,RezendeMW14}, in this work since PAEs have been shown to reach lower reconstruction errors, exhibit better generative properties and, most importantly, reach higher anomaly detection accuracy than their variational counterparts~\citep{PAE}. PAEs also do not require any annealing schemes or fine-tuning during training and therefore alleviate many of the well known problems associated with the optimization of Variational Autoencoders~\citep{FixElbo,Hoffman2016ELBO}.

The second stage of the PAE training is a normalizing flow~\citep{RippelAdams13,DinhKB14,DinhSB16,KingmaD18,ffjord18}, a neural density estimator, which is trained on the encoded data. A normalizing flow (NF) is a bijective mapping $b_\varphi$, parameterized by a neural network with trainable parameters $\varphi$. The NF maps the training data into a space where it follows a tractable target distribution, $p_t$. We make the common choice of choosing a standard normal distribution as target distribution in this work. Normalizing flows are designed to have a computationally tractable Jacobian determinant, which enables the estimation of the log probability density at a data point $z=f_\phi(x)$ through,
\begin{equation}
\log p (z) \approx \log p_\varphi(z) = \log p_t (b_\varphi^{-1}(z)) + \log \det \left[\frac{\partial b^{-1}_\varphi(z)}{\partial z}\right].
\end{equation}
The training objective of normalizing flows is to maximize the estimated average log probability under the model. The loss function is 
\begin{equation}
\mathcal{L}_{\mathrm{NF}}(\varphi)= \underset{x \sim p(x)}{\mathbb{E}}\left[-\log p_\varphi[f_\phi(x)]\right],
\end{equation}
where only the NF parameters $\varphi$ are optimized in the neural network training. Normalizing flows are powerful density estimator that outperform  Kernel Density Estimators in their generalization properties.

The normalizing flow in the PAE maps the irregular encoded distribution into a standard normal distribution. This two-step process ensures that the PAE first identifies an optimal compression before it Gaussianizes the latent space. A Variational Autoencoder is trained to achieve a similar objective, but has to balance reconstruction quality and the regularity of the latent space at the same time. This is known to lead to suboptimal results in practice. 

We illustrate the PAE 2-stage training process in Fig.~\ref{fig:1}.

A probabilistic autoencoder is a generative model. Realistic artificial data samples can be generated by sampling from the Gaussian target distribution of the normalizing flow and passing the samples through both the NF bijector and AE decoder subsequently. 
However, we do not make use of the generative functionality of the PAE in this work. Instead we leverage its anomaly detection functionality. The combination of data compression and density estimation enables accurate out-of-distribution detection based on the density estimate in latent space.

\subsection{Modified PAE for identifying anomalous galaxy spectra}
We use the probabilistic autoencoder to learn the probability distribution of galaxy spectra from training data. We then evaluate the probability of new data points under the PAE model and label data points with low probability as outliers. Galaxy spectra are noisy, incomplete and rich in diversity, which creates the need for a more elaborate anomaly detection pipeline. We describe this PAE-based pipeline in the following sections.

\subsubsection{Denoising and inpainting}
\label{sec:denoise_inpaint}
Given the estimated noise levels from the SDSS-pipeline and the known mask for each spectrum, we modify the autoencoder loss function from mean squared error to $\chi^2$. The $\chi^2$-loss accounts for both pixel-wise noise with variance $\sigma_i$ and binary mask $M_i$,
\begin{equation}
\mathcal{L}_{AE1}(\phi_1,\psi_1) = \underset{x \sim p(x)}{\mathbb{E}}\left[ \sum_{i=0}^{D-1} M_i \frac{\left(x_i-g_{\psi_1}[f_{\phi_1}(x)]_i\right)^2}{\sigma_i^2}\right].
\label{eq:AE1}
\end{equation}
Training on this objective encourages reconstructions that are equal to the maximum likelihood uncorrupted spectrum.
This means that the reconstructions of a $\chi^2$-trained autoencoder  are inpainted and denoised.
\begin{figure}
\centering{
\includegraphics[width=0.45\textwidth]{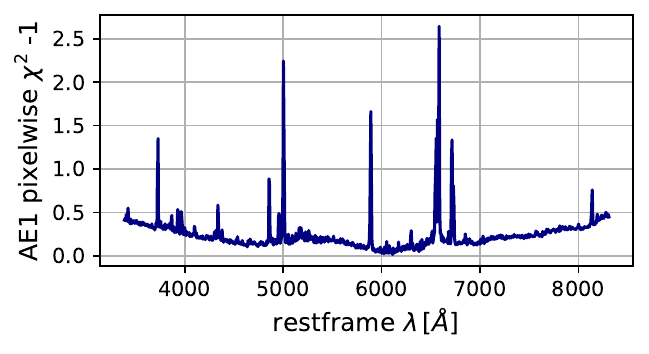}}
\caption{\label{fig:3} Average pixelwise $\chi^2-1$ of the first autoencoder measured on the test set. The first autoencoder is trained on $\chi^2$ loss.}
\end{figure}
\begin{figure}
\centering{
\includegraphics[width=0.45\textwidth]{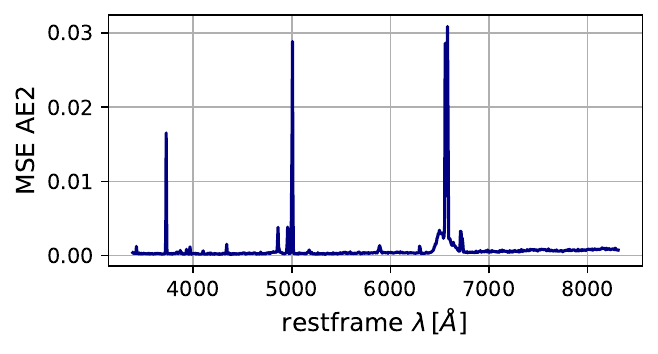}}
\caption{\label{fig:4} The pixelwise mean squared reconstruction error of the second autoencoder, i.e. the mean squared difference between its inputs and outputs. The reconstruction error is extremely small, showing that we sacrifice close to no information about the spectra by adding this second auto-encoding step.}
\end{figure}
We show examples of the input and output to the autoencoder in Figure~\ref{fig:2} and the average pixelwise $\chi^2-1$ in Figure~\ref{fig:3}. 

We find that the mean $\chi^2$ is close to unity near the middle of the spectrum and shows departures 
toward the edges and at the expected location of emission lines. These features are independent of the latent space dimensionality and neural network architecture we choose.  This is not
surprising given observational effects such as variations in throughput, detector sensitivity 
or sky background in particular toward the edges. Furthermore, the SDSS survey team documented 
10-20\% error on the error due to not accounting for the covariance between adjacent spectral 
bins\footnote{\url{https://live-sdss4org-dr16.pantheonsite.io/spectro/caveats/\#perfect}}.
While the wavelength-dependent $\chi^2$ indicates that an optimal training
would require a better description of the error covariance,
particularly in spectral regions with low signal
or high background, we opt to proceed with the published error model.

Upon application of the above autoencoder, we found the
set of extreme outliers to be dominated by galaxies with large
masked fractions (P22; their Figure~12). This effect
is mitigated by inpainting a model prediction
over the masked regions. For instance, \citet{2004AJ....128..585Y, 2020AJ....160...45P} 
used an iterative PCA for the inpainting whereas we
inpaint with the denoised spectra, $x'$ obtained with our first autoencoder.
The resulting inpainted spectra are used to train a second autoencoder of the same architecture and use the encoded distribution of this second autoencoder in the anomaly detection. We train the second autoencoder with a mean-squared error (MSE) loss as objective function,
\begin{equation}
    \mathcal{L}_{\mathrm{AE2}}(\phi_2,\psi_2)=\underset{x \sim p(x)}{\mathbb{E}}\left[\frac{1}{D}\sum_{i=0}^{D-1}\left(x'_i-g_{{\psi_2}}[f_{\phi_2}(x')]_i\right)^2\right].
    \label{eq:AE2}
\end{equation}
The training of the second AE is initialized with the weights from the first AE. We find in our experiments that the pixelwise mean squared error of this autoencoder is extremely low (Figure~\ref{fig:4}), demonstrating that practically no information about the spectra is lost in this additional step. We note that other deep learning approaches that motivate robustness to data corruptions, such as contrastive learning might be employed instead of two autoencoders. We choose AEs in order to evaluate the reconstruction error as an additional anomaly metric. 
\subsubsection{Density Estimation}
We then move on to fit a normalizing flow to the encoded distribution of the second AE. A plethora of normalizing flow models have been proposed in the literature. Here, we use a Sliced Iterative Normalizing Flow (SINF)~\citep{DBLP:conf/icml/DaiS21}. The concept behind this flow is to identify directions in which the difference between the current and target marginal distributions are maximal and fit transformations to reduce this difference one direction at a time. SINF achieves state-of-the-art results on standard machine learning datasets and requires little to no hyperparameter tuning. 
\subsubsection{Conditional Density by Spectral Class}
\begin{figure*}
\includegraphics[width=0.98\textwidth]{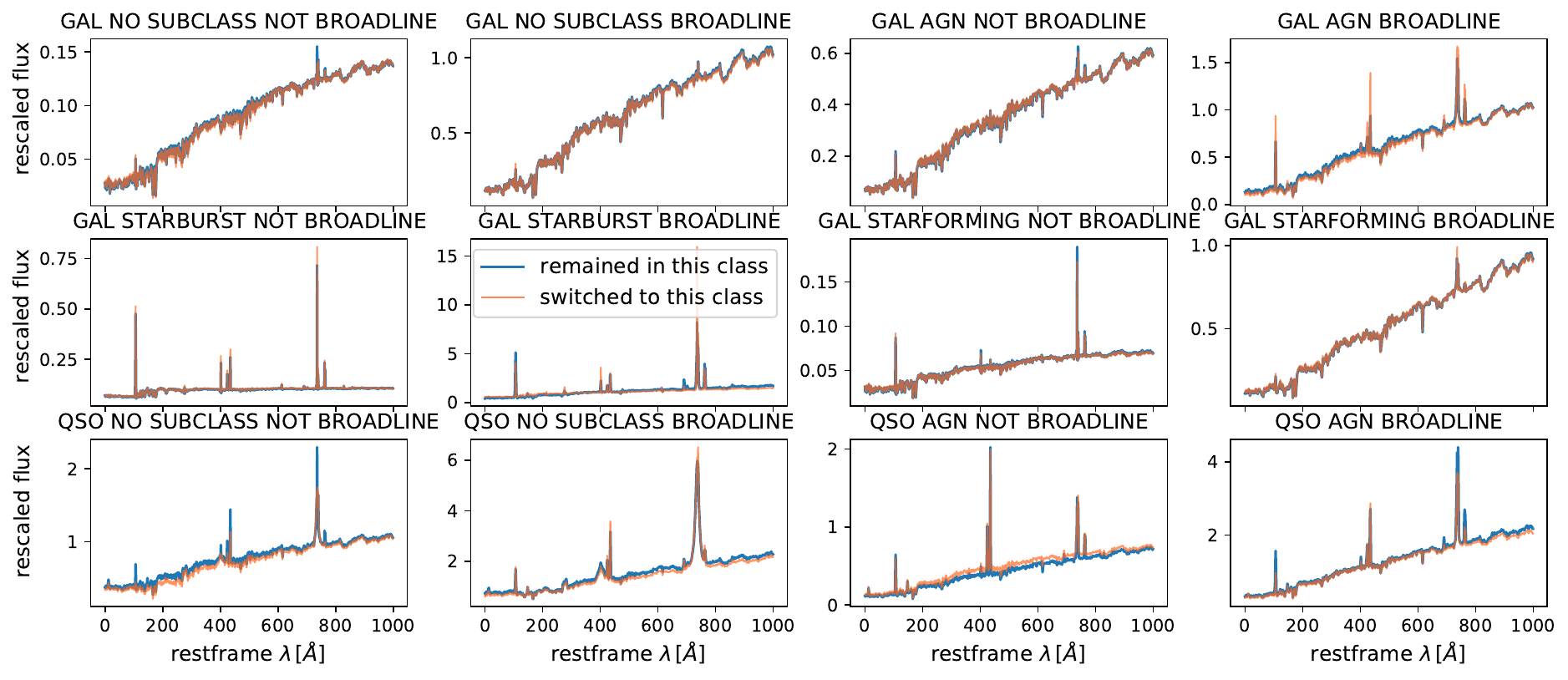}
\caption{\label{fig:5} Examples of spectra that switch class in the relabeling process. For each class (each panel) we show a spectrum that was newly assigned to this class in red together with a similar spectrum that remained in the class (blue).}
\end{figure*}
Galaxy spectra form a very diverse data set. To better organize the results of our algorithm we separate spectra by their spectral class and perform anomaly detection conditional on the class. We start by using pre-defined classes assigned by the SDSS pipeline and provided in the \verb|CLASS| and \verb|SUBCLASS| parameters. The class parameter separates the data set into quasars and galaxies. Galaxies can be assigned an additional subclass parameter, which is determined by their line ratios and widths. The subclass parameters are \verb|STARFORMING|, \verb|STARBURST| and \verb|AGN|\footnote{Active galactic nucleus}. An additional parameter \verb|BROADLINE| is used to label spectra with emission lines that surpass a certain width threshold \footnote{The classification scheme is described in detail on the \href{https://www.sdss.org/dr17/spectro/catalogs/\#\#objects}{SDSS website}.}.  
We use these class labels (first and second row in Table~\ref{tab:labels}) to train a conditional density estimator, which estimates the conditional probability of each encoded data point, $p(z|\mathrm{class})$. The bounds that determine the SDSS classes are sharp cutoffs in a continuous space. For our application, the exact definition of the classes are irrelevant, instead we are interested in grouping spectra by similarity. We therefore define new classes by reversing the process after training the density estimator on the pipeline assigned labels: We use the trained density estimator to determine the most likely class of each spectrum under the model,
\begin{equation}
\mathrm{class_{max}} = \underset{\mathrm{class} \in \mathrm{classes}}{\mathrm{argmax}} \left[\log p_\varphi(z|\mathrm{class})\right].
\end{equation}
We find that a very small fraction (0.75\%) of spectra were assigned a different CLASS from galaxy to QSO or vice versa but about $25\%$ of spectra are assigned to a new SUBCLASS in this process. These changes are mostly minor in the sense that the new labels tend to be similar to the original ones. We show examples of spectra that change class (\verb|CLASS|$+$\verb|SUBCLASS|) under this reassignment alongside their nearest neighbor among the spectra that remained in the newly assigned class in Figure~\ref{fig:5}. This demonstrates that the relabeling indeed assigns spectra to new classes by similarity. 

We retrain the density estimator on the new labels. The re-labeling step is not strictly necessary, since we marginalize over all possible classes for our anomaly detection. However, having less `noisy' labels during training helps the density estimator reach lower loss and thus higher accuracy. The number of objects in each class is tabulated along with an assigned galaxy category to broadly split the test sample into quiescent galaxies, star-forming or AGN host (SF/AGN) galaxies and QSOs (Table~\ref{tab:labels}). 

\begin{table}
\centering
\begin{tabular}{llll}
Label     &       Class Label Name            &  Category     &   N   \\ \hline \hline
    0     & GAL NO SUBCLASS NOT BROADLINE     & quiescent     & 60427 \\
    1     &     GAL NO SUBCLASS BROADLINE     & quiescent     & 11991 \\
    2     &         GAL AGN NOT BROADLINE     &    SF/AGN     &  8017 \\
    3     &             GAL AGN BROADLINE     &    SF/AGN     &  1217 \\
    4     &   GAL STARBURST NOT BROADLINE     &    SF/AGN     & 11835 \\
    5     &       GAL STARBURST BROADLINE     &    SF/AGN     &   107 \\
    6     & GAL STARFORMING NOT BROADLINE     &    SF/AGN     & 39319 \\
    7     &     GAL STARFORMING BROADLINE     &    SF/AGN     &  3168 \\
    8     & QSO NO SUBCLASS NOT BROADLINE     &       QSO     &   186 \\
    9     &     QSO NO SUBCLASS BROADLINE     &       QSO     &  1012 \\
   10     &         QSO AGN NOT BROADLINE     &       QSO     &   396 \\
   11     &             QSO AGN BROADLINE     &       QSO     &   253 \\
   12     &   QSO STARBURST NOT BROADLINE     &       QSO     &    74 \\
   13     &       QSO STARBURST BROADLINE     &       QSO     &  1212 \\
   14     & QSO STARFORMING NOT BROADLINE     &       QSO     &    13 \\
   15     &     QSO STARFORMING BROADLINE     &       QSO     &   415 \\
\end{tabular}
\caption{\label{tab:labels} List of labels with the label number, class label name, assigned galaxy category and the number of galaxies from the test set for each class after the relabeling step.}
\end{table}

\subsubsection{Anomaly Score}
Our final anomaly score, $\mathrm{AS}_{PAE}$, is the probability of a data point marginalized over all possible classes
\begin{equation}
\label{eq:score}
    \mathrm{AS}_{PAE} = - \log p_\varphi(z) = - \log \sum_{\mathrm{classes}} p_\varphi(z|\mathrm{class}) p(\mathrm{class}),
\end{equation}
where the prior $p(\mathrm{class})$ is determined by the frequency of each class in the training set (after re-labeling). 
We use the scipy's \citep{2020SciPy-NMeth} {\tt logsumexp} function to perform the marginalization in a numerically stable manner.
\subsection{PAE architecture and training}

\begin{table}
\centering
\begin{tabular}{llll}
Parameter & Value Range & Scale & Value \\ \hline \hline
Network Depth & $[2,5]$ & lin & 2 \\
FC Units & $[\mathrm{Latent\, Size}, \mathrm{Layer\, Input\, Size}]$ & lin & 100,590 \\
Latent Size & $[2,14]$ & lin & 10 \\
Dropout Rate & $[1\times10^{-3},0.5]$ & log & 0 \\ \hline
Initial L-Rate & $[1\times10^{-4},2\times10^{-3}]$ & lin & $7\times 10^{-4}$\\
Final L-Rate & $[5\times10^{-6}$, Initial L-Rate] & log & $1.3\times 10^{-5}$ \\
Decay Steps & $[2000,40000//\mathrm{Batch\, Size}\times10]$ & log & 2300 \\
Optimizer & [Adam, RMSprop, SGD] & - & Adam \\
Batch Size & $[16,256]$ & lin & 33 \\
\end{tabular}
\caption{\label{tab:1} Parameters governing the autoencoder network architecture and training procedure that were optimized in order to achieve low reconstruction error. The values used in the analysis are listed in the last column.}
\end{table}

We choose a standard multilayer perceptron (MLP) architecture for the autoencoder, which we found to reach lower reconstruction error than convolutional architectures, but verified that other architectures result in anomaly scores that are highly correlated with the presented results. The architecture and training parameters are optimized with \verb|OPTUNA|~\citep{optuna_2019}, an automatic hyperparameter optimization software framework for machine learning. We run ca. 1500 training trials over 10 epochs to sample the network performance measured in terms of the validation loss as a function of these parameters. The parameters we vary, the explored ranges and (rounded) values for the best performing model are given in Table~\ref{tab:1}. We build the decoder as an exact mirror of the decoder. The best performing model consists of three layers. We optimized the number of units in the first two layers (of the encoder). Both of them are followed by a LeakyReLU activation function and  Dropout layer. The last layer is a fully connected (FC) layer without activation that maps the output of the previous layers to the desired latent space dimensionality. Tests with convolutional architectures did not result in better reconstruction performance. We use a polynomial decay schedule from an initial learning rate (initial L-Rate) to a final learning rate (final L-rate) over an optimized number of decay steps. 
The second autoencoder stage shares the same architecture as the first one and we use the pre-trained weights from the first stage as initial values. We monitored the loss on the test set during training of the autoencoders and did not observe any overfitting. The density estimation is trained for 50 epochs. We chose the Sliced Iterative Normalizing Flow as a density estimator not only because of its proven performance, but also because it requires practically no hyperparameter tuning. The model simply adds transformation layers until the loss on the test set stops improving. 

\section{Results}
\label{sec:OOD}

The test set is diverse as it includes QSOs, quiescent galaxies, and SF/AGN galaxies (Table~\ref{tab:labels}). 
After briefly investigating the distribution of properties and probability scores for each category of galaxies (\S\ref{sec:galpop}), 
we will focus our analysis on quiescent galaxies because they represent a more homogeneous population as a starting point. 
By focusing on spectra that do not fall into one of the more easily defined categories with strong spectral signatures $-$ such as starburst, AGN, QSO, etc. $-$ we have a chance of uncovering some of the more subtle anomalies that the algorithm picks up and that might be difficult to identify otherwise.

\begin{figure}
\includegraphics[width=0.4\textwidth]{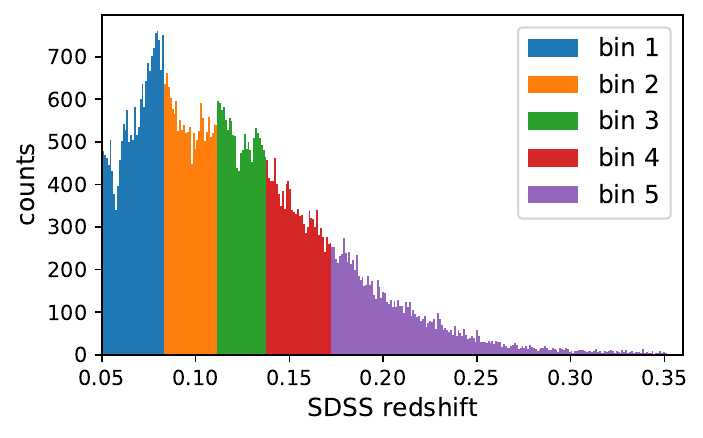}
\caption{\label{fig:6} Redshift distribution of quiescent galaxies from the test sample and definition of redshift bins.}
\end{figure}
To stratify our analysis, we further split the test quiescent galaxy sample into five redshift bins (Figure~\ref{fig:6}) with each bin containing about 20\% of the sample. This ensures that presented anomalies are not biased by redshift dependencies. 

We then use both, the density-based anomaly detection score proposed in this work, $\mathrm{AS}_{PAE}$, as well as the AE reconstruction error to identify the top-eight most outlying spectra in each redshift bin (Tables~\ref{tab:A1} and ~\ref{tab:A2}).
The anomaly detection score is good at identifying rare objects with counterparts in
the training sample, whereas
reconstruction error is good at detecting objects not at all represented in the training sample.
We determine the most likely reason for the anomaly of each spectrum by visual inspection of the spectrum, the corresponding image, and in some cases a derivative-based sensitivity analysis. In the sensitivity analysis we take the derivative of the anomaly score with respect to the (encoded) input spectrum, and use a gradient descent method to track how the spectrum changes as we follow the gradient to higher probabilities. We discuss the most common types of anomalies in the following subsections.  In some cases, a data
or astronomical anomaly also results in a faulty SDSS redshift determination,
exacerbating its outlier score.

Objects are referred to by their SDSS
{\tt (MJD, plate, fiber)}.

\subsection{Galaxy population global properties and probability scores}
\label{sec:galpop}

\begin{figure}
\begin{center}
\includegraphics[width=0.45\textwidth]{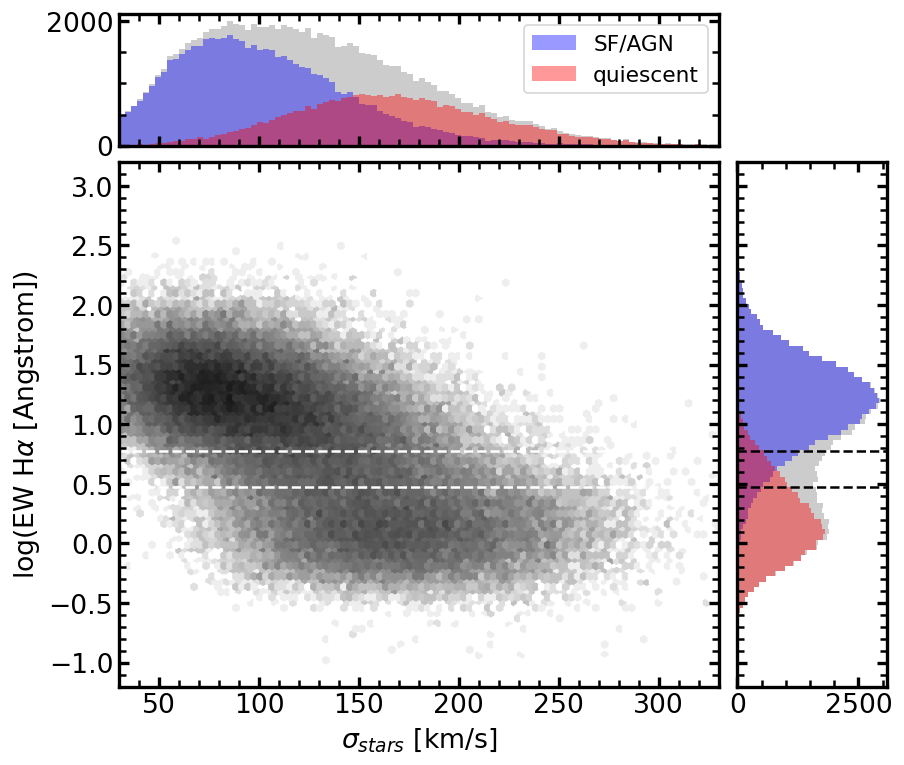}
\caption{\label{fig:ewsig_all} \ha\ equivalent width as a function of the stellar velocity dispersion for the test sample. The top and side panels show the total distribution along each axis (grey) as well as the separate distributions for the quiescent galaxies (red) and SF/AGN galaxies (blue). Horizontal dashed lines correspond to 3~$\AA$ and 6~$\AA$, where there is significant overlap between the SF/AGN and quiescent galaxy populations.}
\end{center}
\end{figure}

To illustrate global properties of the galaxy populations from the test set, we show in Figure~\ref{fig:ewsig_all} their bivariate distribution in terms of \ha\ equivalent widths $-$ a proxy for their specific star formation rate $-$ and $\sigma_{\textrm{stars}}$ $-$ a proxy for their mass. This figure recovers a well known galaxy bimodality between star-forming and quiescent galaxies \citep[e.g.,][]{Strateva+2001,Mateus+2006}. On the one hand, SF/AGN galaxies tend to have high values of EW(\ha) due to ongoing star formation and/or nuclear activity. They also preferentially exhibit low and mid range values of $\sigma_{\textrm{stars}}$. On the other hand, quiescent galaxies preferentially have low values of EW(\ha) consistently with little to no ongoing star formation, and are found to reach high values of $\sigma_{\textrm{stars}}$, consistently with the majority of massive galaxies at low redshift being ``quenched''.

While we expect that low-mass (and therefore low $\sigma_{\textrm{stars}}$) quiescent galaxies exist, they are much more challenging to detect owing to their faint apparent magnitudes. Therefore, the lack of their representation is likely due to survey detection limits and the difficulty in measuring extremely faint emission lines. Similarly, the lower envelope around $EW(\ha)\sim0.5~$\AA\ is due to the emission line detection limit as we impose a signal-to-noise cut of $S/N>2$ in the \ha\ emission line for Figure~\ref{fig:ewsig_all}. 
Of the 72,418 quiescent galaxies in the test sample 33,690 (47\%) pass the \ha\ S/N cut while the remaining 53\% are not detected in \ha\ or missing $\sigma_{\textrm{stars}}$ information (0.2\%) and thus not shown on the figure. In contrast, the bulk of the 67,224 SF/AGN galaxies have an \ha\ detection with only 0.35\% failing the S/N cut. However, we note that 6,486 (9.6\%) SF/AGN galaxies are not shown on Figure~\ref{fig:ewsig_all} due to unresolved (<30 km/s) $\sigma_{\textrm{stars}}$ or due to missing a DR12 value added catalog counterpart (5.6\% and 4\%, respectively). Despite these limitations, it is clear that the SF/AGN and quiescent galaxy populations strongly differ from one another. In fact, the difference in \ha\ EW values would be even more pronounced if we could detect yet fainter emission lines.

\begin{figure}
\begin{center}
\includegraphics[width=0.35\textwidth]{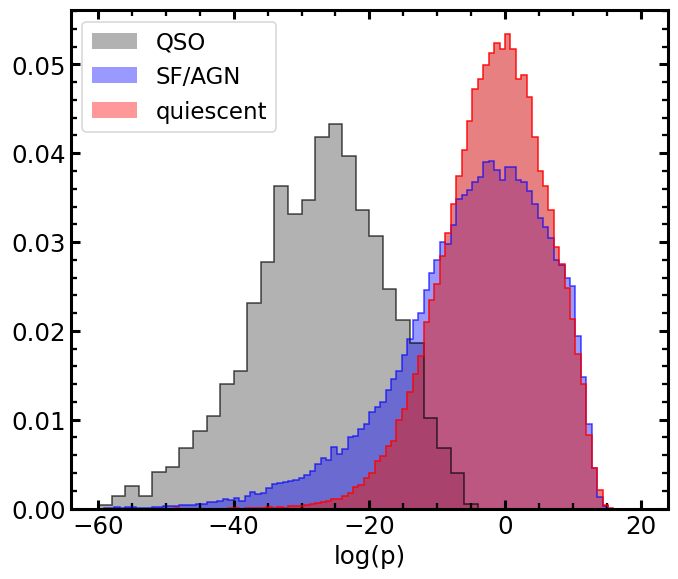}
\caption{\label{fig:logp} Distribution of $\log{p}$ values for QSOs (grey), quiescent galaxies (red) and SF/AGN galaxies (blue) in the test sample.}
\end{center}
\end{figure}

The QSO category is omitted from Figure~\ref{fig:ewsig_all} because $\sigma_{\textrm{stars}}$ is not available for spectra dominated by the power-law continuum from the accretion disk as is often the case for luminous QSOs. This important difference in spectral shape also means that we expect QSO spectra to score lower values of $\log{p}$ relative to the bulk of non-QSO galaxies. To a lesser extent, we expect the SF/AGN galaxies to have a more significant tail of less probable (more anomalous) spectra due to phenomena like starbursts and strong AGN episodes that produce obvious spectral signatures such as enhanced emission line strengths and widths. Indeed, we can observe these trends by comparing the normalized distributions of $\log{p}$ values for the QSOs, SF/AGN galaxies, and quiescent galaxies (Figure~\ref{fig:logp}). As expected, the quiescent galaxies (in red) have the narrowest peak with the least prominent tail toward low probability scores (i.e., high anomaly scores). This motivates our choice to focus on the quiescent galaxies in the remainder of this paper.

\subsection{Identified anomalies}
\label{sec:anomaly}
\begin{figure*}
\includegraphics[width=0.4\textwidth]{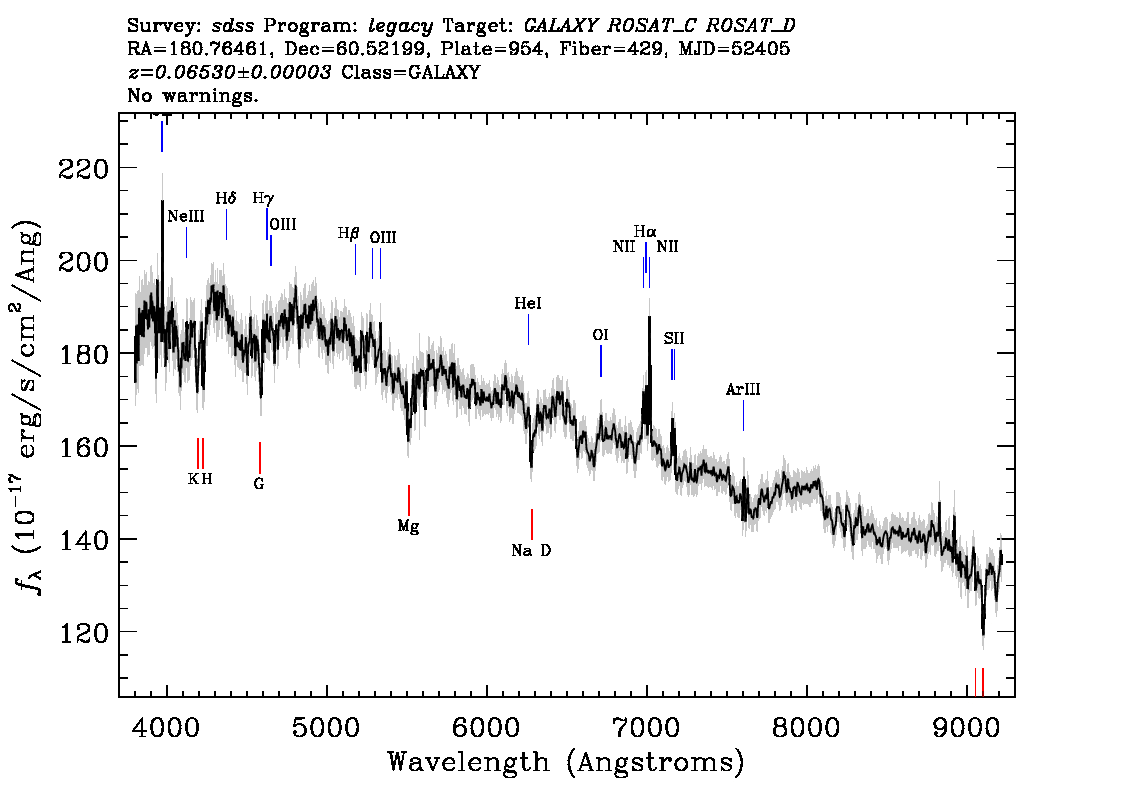}
\includegraphics[width=0.4\textwidth]{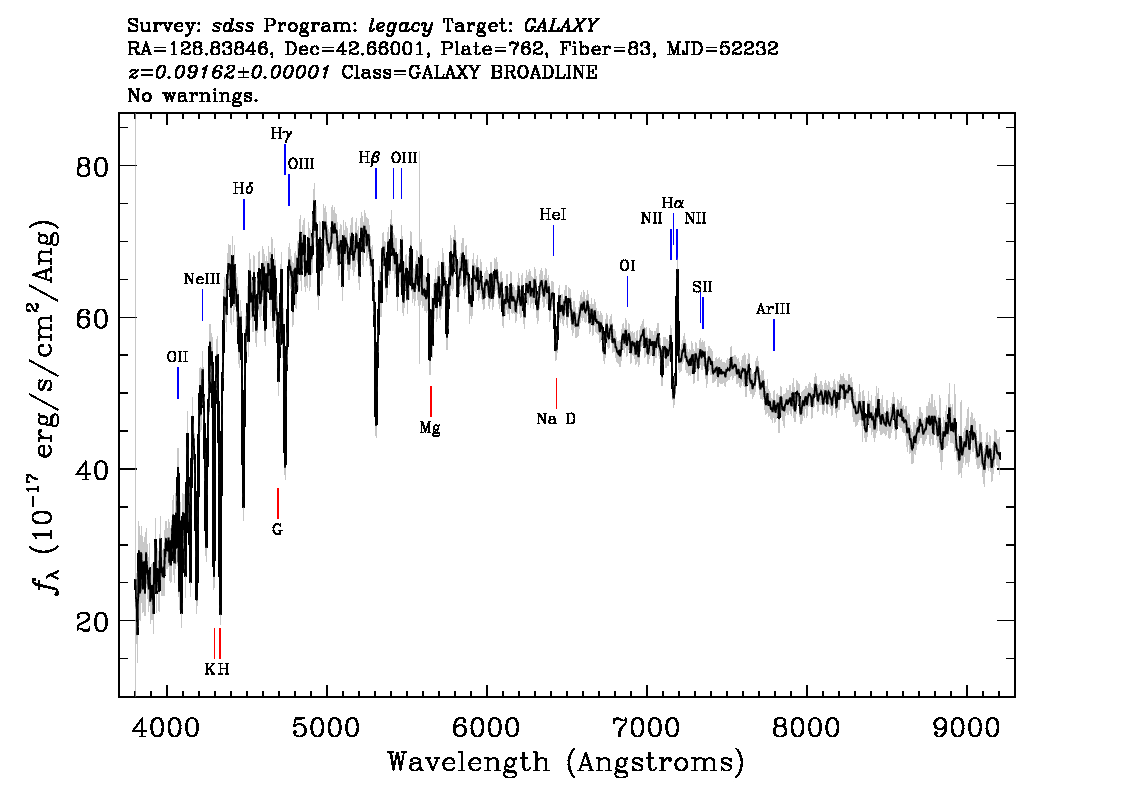}
\includegraphics[width=0.4\textwidth]{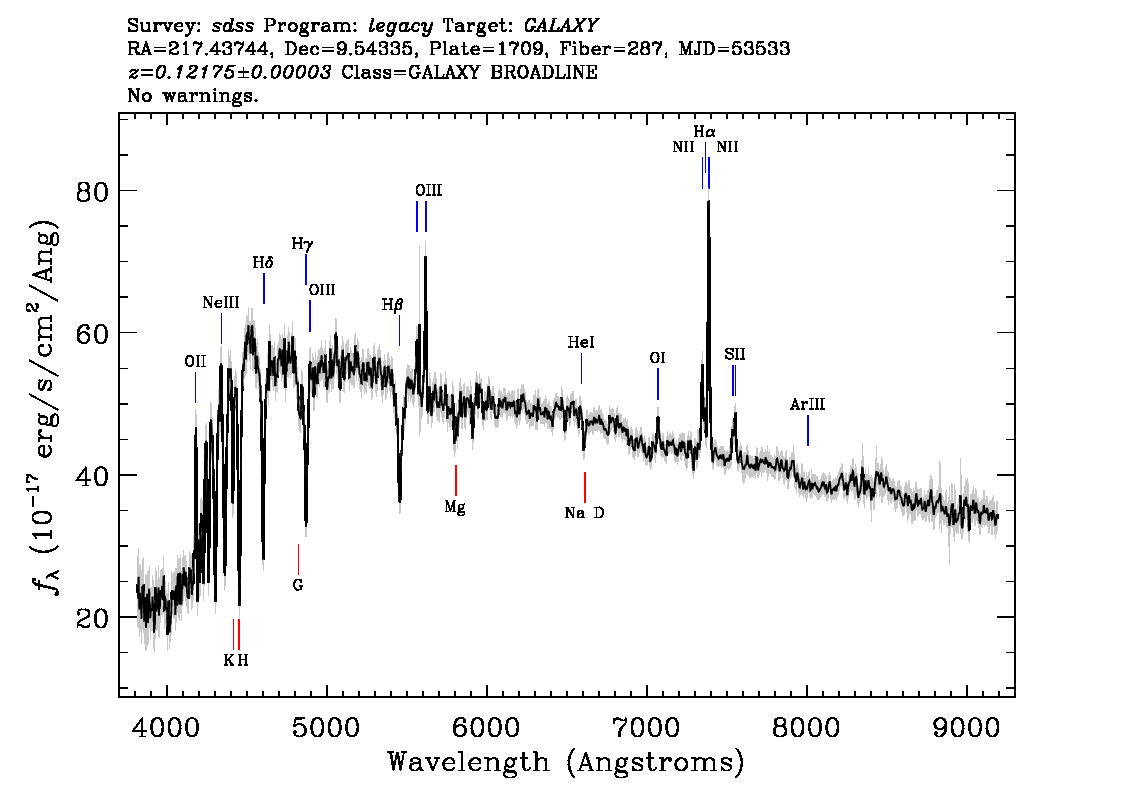}
\includegraphics[width=0.4\textwidth]{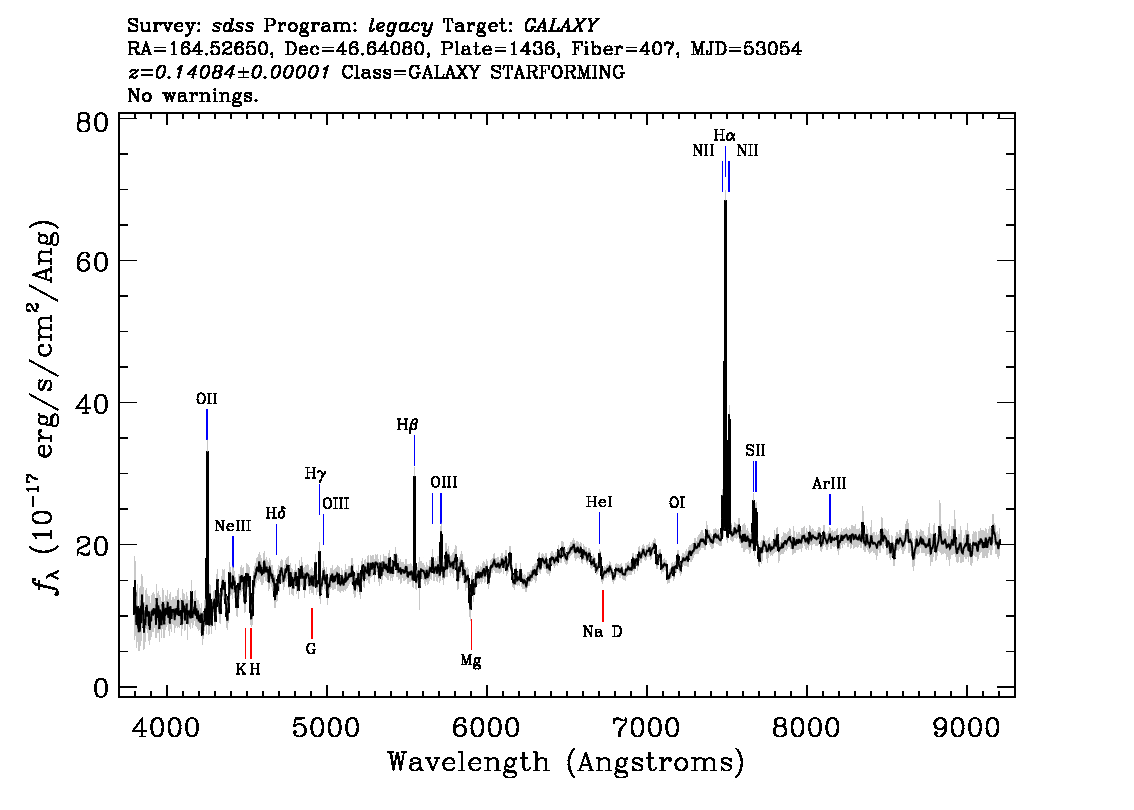}
\includegraphics[width=0.4\textwidth]{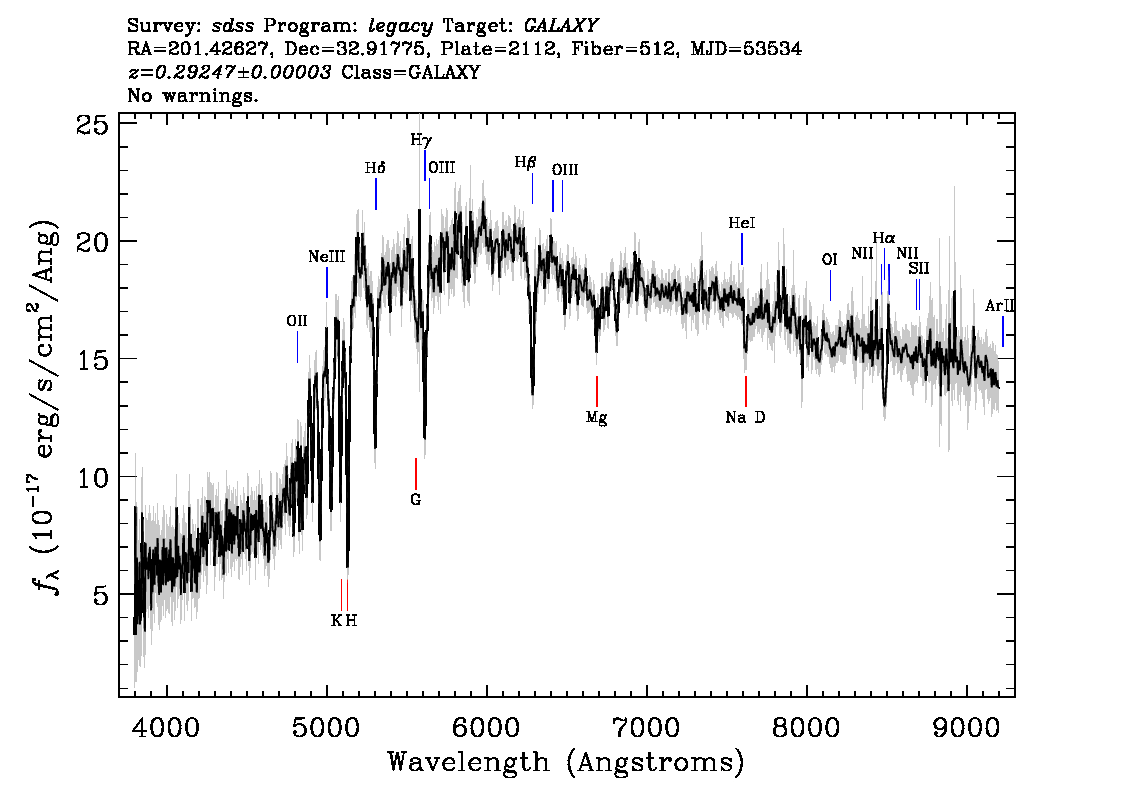}
\caption{\label{fig:topanomoly} The most anomolous spectrum for each of the five redshift bins, with increasing redshift going from top left to bottom right. Each panel, generated with the SDSS SkyServer, shows the original spectrum in gray, a smoothed version in black, and labels at the expected location of emission and absorption lines in blue and red, respectively. Information on each target is listed above its corresponding spectrum.}
\end{figure*}

The anomaly score is used to identify those
spectra which, after compression into
latent-parameter space, have parameter
values that are the least probable.
The eight most anomalous spectra
for the five redshift bins 
are listed in  Table~\ref{tab:A1} and
the SDSS 
spectrum plot \citep{2000AJ....120.1579Y,2002cs........2013S}
of highest outlier per redshift bin is shown in
Figure~\ref{fig:topanomoly}.
Each panel was generated using the SDSS SkyServer\footnote{https://skyserver.sdss.org/dr16/en/tools/explore/summary.aspx} 
and includes information on the corresponding target.

The original spectrum, its
reconstruction, and the closest
training spectrum along the path in latent
parameter space
that  maximizes $\log{p}$ ascent
were visually inspected
for each of the forty outliers.
The anomalies can be categorized
as follows.  In some cases 
the reconstructions are poor, indeed 
six objects that are extreme anomalies also
appear in the extreme
reconstruction outlier list to be introduced in
\S\ref{sec:reconstruction};
it is not surprising that their
latent parameters are improbable.
In other cases, the
reconstruction is good,
the slope of ascent in
$\log{p}$ with
respect to the latent parameters is large, and it is
clear by the difference between
the spectra along
the ascent which features
are ``anomalous''.
In most cases, the reconstruction is good but the slope of 
$\log{p}$ is shallow, and produces only subtle differences between the spectra.
These objects lie at the tail of
a continuous distribution.

In the following subsections,
we identify the physical causes
that make the outliers
outliers.  In a few cases, physical effects give
rise to an incorrect redshift, which in turn
exacerbates the outlier appearance of these sources.

\subsubsection{AGN/QSO}

There are two objects that are visually clear
outliers that can be assigned as AGN.
Object (52405, 954, 429)
is the top outlier in
redshift bin 1 and is shown
in Figure~\ref{fig:topanomoly}.
Its spectrum is dominated by a strong blue continuum with comparatively weak emission lines relative to typical QSOs. 
This object was targeted by SDSS as a ROSAT source \citep{2016A&A...588A.103B}
that is bright and/or moderately blue ({\tt ROSAT\_C})
and also bright enough for follow-up spectroscopy ({\tt ROSAT\_D}).
It has been classified as a BL Lacertae object \citep{Anderson+2007,Plotkin+2008}, which is a rare QSO sub-type. This classification is reported in the SIMBAD astronomical database 
\citep{2000A&AS..143....9W}, and differs from both the SDSS pipeline and PAE classification as a GALAXY. 
The erroneous spectral types and large anomaly score likely all arise from the unusual spectral shape which appears somewhat intermediate between a featureless BL Lac blazar (BZB) spectrum and a BL Lac Galaxy-dominated (BZG) spectrum as defined by \citet{deMenezes+2019} and lacks strong broad QSO lines.

\begin{figure*}
\includegraphics[width=0.4\textwidth]{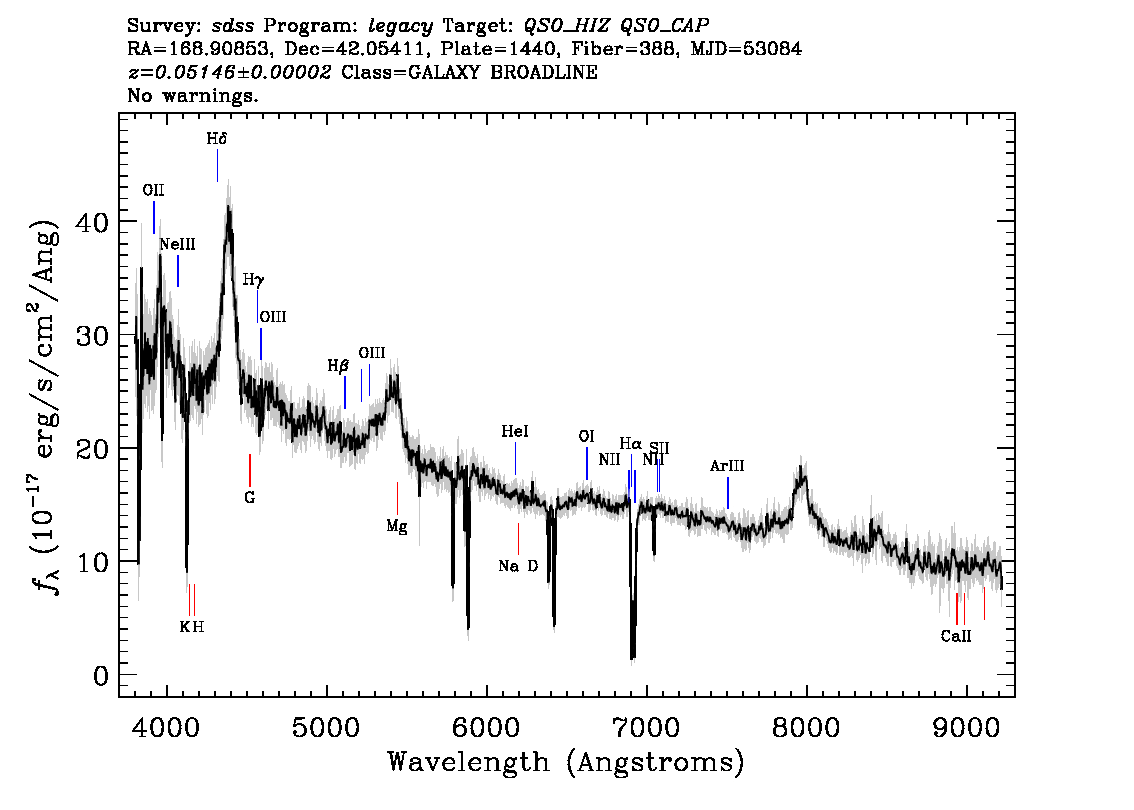}
\includegraphics[width=0.4\textwidth]{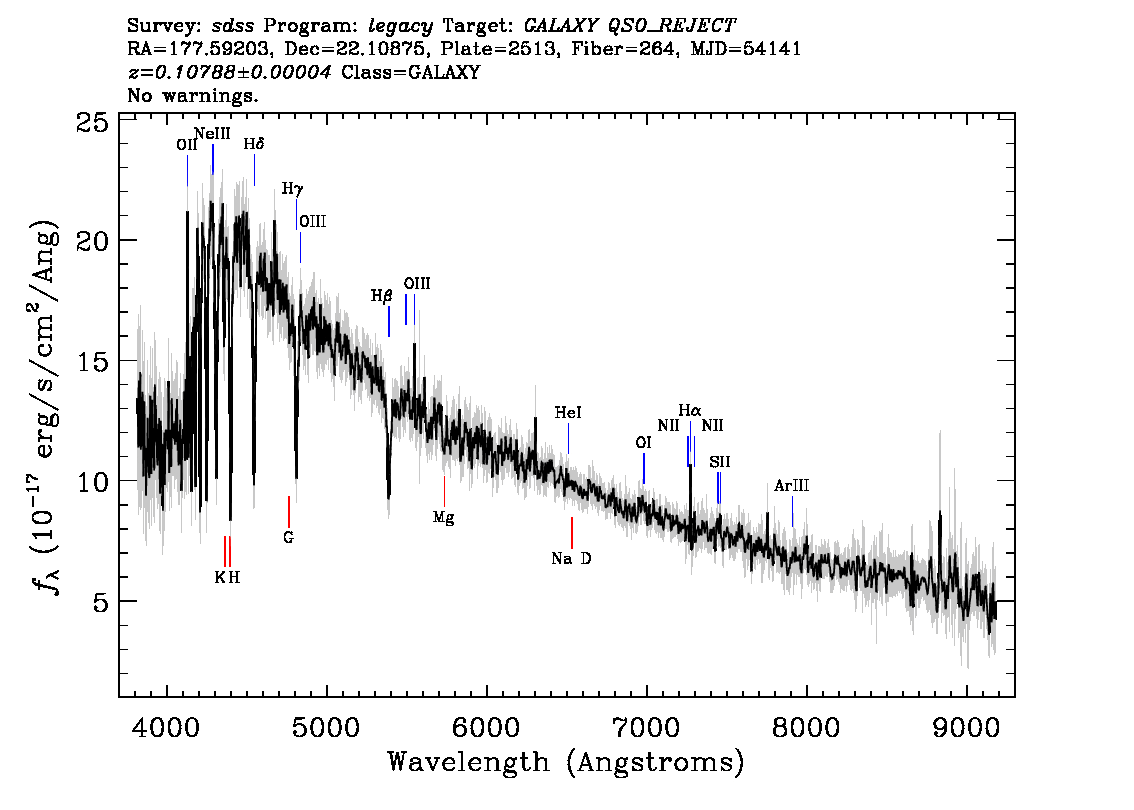}
\caption{\label{fig:other_outliers} Left: Spectrum of
object (53084, 388, 1440), an anomaly
identified by our pipeline, which is
a quasar assigned to an incorrect redshift
and incorrect spectral type by the SDSS pipeline.
Right: Spectrum of
object (54141, 2513, 264), the one
example that is unlike the others in the
\EA{} anomaly sample, due to its
extremely blue color and its lack of
prominent [NII] lines.  
This object was targeted by SDSS
as a {\tt GALAXY}; it was also
tagged as being in an area of QSO color-color
space with a high contamination rate, hence the
{\tt QSO\_REJECT} label.}
\end{figure*}

The SDSS redshift of the
second clear AGN/QSO outlier is incorrect.
Object (53084, 388, 1440), whose spectrum
is shown in Figure~\ref{fig:other_outliers}, is
identified as a broadline
galaxy at $z=0.05$ by SDSS though it was targeted
as and 
is in fact a quasar at $z=1.84$.
The PAE reconstruction of this object
is also poor. 
This is expected as such a catastrophic redshift failure would not be represented in the training dataset, and our approach does not vary the redshift and assumes correct rest-frame wavelengths. 

More AGNs and quasars among our outliers that share
 distinguishing features are described in subsequent subsections.

\subsubsection{\EA{} Galaxies}
\label{e&a:sec}
The majority of the outliers are visually
classified as \EA{} galaxies, whose
spectra exhibit both old (elliptical)
and young (represented by A stars) stellar
populations \citep{1983ApJ...270....7D}.  They are
characterized by
having deep Balmer absorption lines but no significant [O II] emission,
indicative of being post-starburst \citep[see review by][and references therein]{French2021}.
Objects (53533, 1709, 287) 
and
(53534, 2112, 512) shown in Figure~\ref{fig:topanomoly} are examples of
this type of outlier.

Almost all
of these objects exhibit the presence
of \nii\ and weak to absent \ha,
and so are classified as either AGN or galaxy with a low-ionization nuclear emission-line region \citep[LINER;][]{Heckman1980} based
on their spectral features and/or position on the BPT emission-line diagnostic diagram \citep{1981PASP...93....5B}.
Three of the \EA{} galaxies, including (53533, 1709, 287), also have
a prominent [OIII] line, putting them into the AGN class.
Note that \citet{2017A&A...597A.134M}
find that only 2-3\% of \EA{} galaxies
are found to host AGNs using conventional search criteria.

From among this class of outliers, one stands out
as being particularly unlike the others.
Object (54141, 2513, 264) is extremely blue,
dominated by the spectra
of OBA-type stars, and does not have
\nii\ emission.  The spectrum
of this object is shown in
Figure~\ref{fig:other_outliers}.
The optical image from the Legacy Surveys \citep[LS;][]{Dey+2019} Sky Viewer\footnote{Accessed through the URL \url{https://www.legacysurvey.org/viewer?ra={RA}&dec={Dec}&zoom=16}.} 
shows the galaxy to be
blue, diffuse, and asymmetric, indicating
possible tidal interaction.

We conclude that \EA{} galaxies with evidence
of AGN activity through their significant \nii\ emission
are relatively rare in our sample and these spectra
that make the top outlier list represent
the extreme tail of their distribution.
Indeed, these are the rare galaxies
that bridge the dominant populations of
red inactive galaxies and blue galaxies
with significant star formation \citep{2017A&A...597A.134M}.

\subsubsection{\nii\ emission/LINERs}
Most of the anomaly sample
have a high \nii/\ha\ ratio, with
almost all of those exhibiting weak to no
\oiii, \ha, nor \hb\ emission.
Based on these features, these anomalies
are identified as LINERs, as distinguished
from star-forming
galaxies and 
normal AGNs, based on their inferred 
position in the 
BPT diagram \citep{1981PASP...93....5B}.

In details, different mechanisms have been proposed to 
explain LINER-like spectral signatures such as weak AGNs \citep{Ho2008} or retired galaxies with spectra dominated by an aging stellar populations with gas ionized by low-mass stars \citep{CidFernandes2011}. Similarly, \citet{Agostino2021} recently presented an approach to distinguish two sub-populations differing in the 
hardness of the ionization field (dubbed 
{\it soft} and {\it hard} LINERs). While a detailed investigation is beyond 
the scope of this exploratory work, the identification of LINER-like spectra as PAE anomalies could motivate future effort.

\subsubsection{Neighboring Contaminating Sources}
\label{sec:contamination1}
Inspection of 
LS DR9 optical imaging
using the LS Sky Viewer
showed four cases of a
secondary source within the fiber aperture placed at the primary target.
An example is
(53054, 1436, 407), shown
in Figure~\ref{fig:topanomoly}, which looks like an M star overlapping
a galaxy.  This object is also
selected as having a poor reconstruction, as seen in Figure~\ref{fig:recons_examples}:
the features
of the M star are not accommodated by the PAE model.
The odd occurrence of multiple sources entering
a fiber has already been documented in
\citet{2012AJ....144..144B}, who present example
contaminated spectra from the BOSS Survey.

In addition, there were a number of cases of possibly overlapping
sources. Spectroscopically, the
visual characteristic for these anomalies is
a slight tilt in the observed continuum
color relative to the reconstructed
continuum color.

\subsubsection{Supernovae}
\label{sec:sn1}
One outlier is SN2001km,
a Type~Ia supernova discovered
by \citet{Madgwick_2003}.

There are other supernovae in our
sample previously
identified using template-based (not machine learning) algorithms \citep{Madgwick_2003,2013MNRAS.430.1746G}.  The distribution
of their anomaly scores is poor
relative to the general population, 
but are not isolated in the tail.
The PAE outlier criterion in this work was not tailored to identify
known SNe within SDSS spectra, but the procedure could be adapted and calibrated to be more sensitive to a specific type of anomaly such as supernova. Curent machine-learning
methods tailored for
supernova detection and classification in spectra \cite[e.g.][]{2019PASP..131k8002M, 2021GCN.30923....1P}
are based on supervised learning.

\subsubsection{Local low surface brightness galaxies}

Object (53520, 1657, 126) 
is one of four sources with similar and 
poor PAE reconstructions,
which we show in \S\ref{recon-ston:sec} to be local ($z<0.1$)  low surface
brightness galaxies.
Of these four, only this one had a top-8 anomaly score.
Note that some other sources with
poor PAE reconstructions that have
been identified
as QSO, supernova, or overlapping sources,
also have top-8 anomaly scores.  Generally,
however,
objects with extreme reconstruction
errors do not necessarily find themselves in the top anomaly list.

\subsubsection{Externally classified Active Galaxies}
In this subsection we depart from discussing
anomalous features identified within
the SDSS spectra, but
rather refer to AGN classifications
reported by SIMBAD based on X-ray data.

\begin{figure*}
\includegraphics[width=0.4\textwidth]{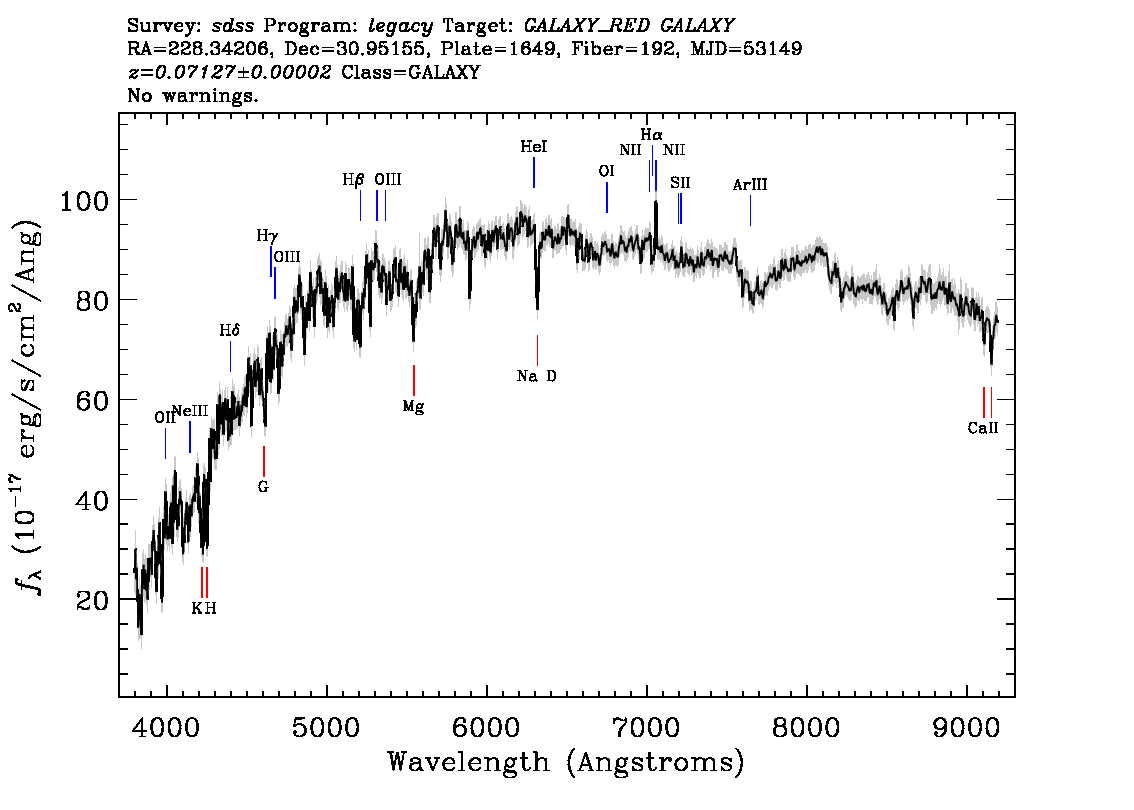}
\caption{\label{fig:B} Spectrum of
object (53149,1649,192) from the SDSS SkyServer.
The source is labeled as a LINER
in the SIMBAD database and the spectrum is that of an
elliptical galaxy.  Note the presence of \nii\ emission and
the absence of \ha.}
\end{figure*}

Seventeen outliers
are identified 
in SIMBAD
as LINER, AGN or QSO.
Of these, slightly
more than half are
``\EA{}'' galaxies.
The remaining ``AGN''
outliers are weak-line radio galaxies
\citep{1979MNRAS.188..111H, 1998MNRAS.298.1035T, 2010A&A...509A...6B},
classified
as radio-loud AGN 
whose optical
spectra  the appearance of
an old elliptical galaxy
and no or weak \oiii\ emission.
An example
of such a spectrum
is shown in Figure~\ref{fig:B}.
All of the weak-line radio
galaxies exhibit \nii\ emission and
no \ha, to which
we attribute to a LINER
rather than an AGN classification.

\subsubsection{Comparison with Previous Work}
\label{sec:baron}

\citet{2017MNRAS.465.4530B} [B17]
performed outlier detection on SDSS data.
They trained a random forest labeling
real data and synthetic spectra drawn 
from the per-feature distribution of the
real data and no covariance between features.  Outliers
were then identified in a metric space based on
distance between their nodes in the forest.

The samples in which we 
and [B17] search for outliers are
significantly different.  We restrict our redshift range, use a higher data resolution,
and divide the spectra into training and test
samples.  In this article we focus on the subset of galaxies that
are not classified as
{\tt STARFORMING}, {\tt STARBURST}, or
{\tt AGN},
whereas [B17]'s approach is particularly sensitive to sharp emission lines that would often put
a galaxy in one of the above classes.

Considering these differences, the classes of outlier
identified in both works are identical.
One class are \EA{} Galaxies with
strong Balmer absorption, both with and without
ongoing star formation.
[B17] find a class
of outliers in the BPT diagram,
at the edges of both [OIII]/H$\beta$
and [NII]/H$\alpha$.  Our
outliers rarely exhibit
either [OIII] or H$\beta$ emission
(a large fraction having  Balmer absorption) but we have identified
high [NII]/H$\alpha$, which we
associate with AGN.  Strong
NaID absorption appears in many of the outliers we associate with AGNs.
Other common outlier classes
are SNe, blends, calibration 
errors.  Our extremely red outliers
have been associated with a red
background source seen in imaging.

\citet{2020AJ....160...45P} used
VAE as a method to reduce the dimensionality of SDSS spectra.
The focus was on
the quality of their reconstruction
and identifying tracks in their latent
space that traverse astrophysical
classes of objects.
Their top ten outliers are explained
as stars that were erroneously classified
as galaxies, low SNR, contaminating
neighbor, missing data, and poor calibration.

\citet[][hereafter P22]{Pat_2022} used the same training and test sets as this work 
and employed a similar method except that they included only one round of AE 
therefore skipping the in-painting step. That earlier work also did 
not employ labels for conditional probability estimation, and was thus more
subject to probability density scores being affected by the relative importance 
of QSO and galaxy populations. As a result, the most outlier spectra tended to 
be biased toward cases with a higher fraction of masked pixels and/or 
QSO spectra (see their Figure~12). While that work was not focused on identifying 
outliers, noting these behaviors has helped improve our choices for the 
present experiment aiming at finding outlier spectra among subsets of galaxies 
(specifically quiescent galaxies in this work).

\begin{figure*}
\begin{center}
\includegraphics[width=0.95\textwidth]{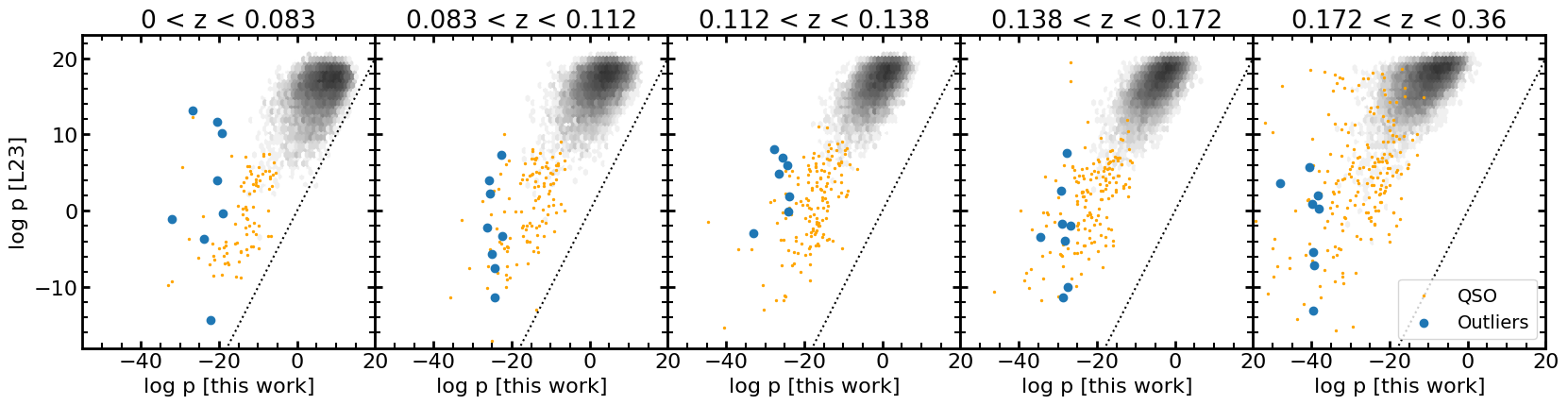}
\caption{\label{fig:logp_comp} Comparison of the $\log{p}$ values from L23 and from 
this work for the overlapping sample of 71,703 quiescent galaxies (grey shaded bivariate 
distribution) split into the five redshift bins used in this work. The redshift ranges 
increase from left to right as labeled. 
We show the outliers based on the lowest anomaly scores from this work (blue filled circles) 
and the 771 overlapping QSO spectra (orange dots). The black dotted line corresponds 
to the one-to-one relationship.
}
\end{center}
\end{figure*}

More recently, \citet[][hereafter L23]{Liang+2023} presented an overall similar 
study to that of P22 and to this work. Despite the general similarity, 
two notable differences lead to slightly distinct and complementary results. 
These authors used the SPENDER neural network architecture \citep{2022arXiv221107890M} 
to train a model of SDSS spectra and inspect outliers. The first important difference 
is that L23 added a loss term to account for redshift variations 
whereas we opted to split the sample into redshift bins. The second difference 
is that we employed a conditional density estimation based on labels whereas L23 
did not use labels $-$ more similarly to P22. To compare the results 
we combined the catalog from L23 with our test set, finding 132,257 common entries 
among our test set of 139,642 (95\%). In particular, we focus on the subset 
of 71,703 quiescent galaxies and show a comparison of the $\log{p}$ values from L23 
and from this work in Figure~\ref{fig:logp_comp}. Each panel corresponds to a 
redshift bin increasing from left to right as labeled.

The outliers based on their anomaly scores from this work are shown with blue circles 
and they tend to also be outliers according to the L23 model. This implies that the two  
trained models are largely consistent in their outlier detection. We also note 
differences that we attribute to the distinct decisions described earlier. 
Namely, the difference in using a loss term to correct for redshift (L23) versus 
splitting the sample into redshift bins (this work) manifests itself as a redshift 
dependent offset when comparing directly $\log{p}$ values (one-on-one dotted line). 
The second difference is that the use (or absence) of conditional probability results 
in the QSO (small orange dots) being more heavily biased toward a tail of outliers 
in the L23 model compared to this model. There are advantages and disadvantages to each 
set of decisions, which ultimately needs to be tailored to the science question at hand.

\subsection{Most outlying reconstruction error}
\label{sec:reconstruction}

We define reconstruction error as the masked and inverse-noise weighted L2-distance between the original input spectrum and the output of the second decoder
\begin{equation}
RE_{PAE} = \sum_{i=0}^{D-1} M_i \frac{\left(x_i-g_{\psi_2}[f_{\phi_2}(x')]_i\right)^2}{\sigma_i^2}.
\end{equation}
The eight  spectra with highest reconstruction error 
for the five redshift bins were identified
and visually inspected. 
These spectra are given in Table~\ref{tab:A2}.
In this subsection, we discuss
sources of reconstruction error in these
forty spectra that were identified by humans.
We do not plot the largest
outlier for each redshift
bin as was done for anomaly detection, as
they do not fairly represent cases of interest.

\subsubsection{Airglow}

Four outliers occur on Plate/MJD
362/51999
that exhibit [OI] airglow emission lines
at observer-frame 6302, 6365\AA.
Of these, three are the largest
outlier in their respective redshift bin.
Inspection of other spectra not in our
top 40 list but
taken in this exposure exhibit the same lines. 
\subsubsection{Contaminating Sources}
Images of eighteen of the forty outliers
exhibit the clear overlap of distinct
astronomical sources.
While the spectral features of the dominant source upon which the fiber is centered is well reconstructed by the PAE, the continuum is not.

One example already noted in \S\ref{sec:contamination1}
is object
(53054, 1436, 407), an M star overlapping
a galaxy.  Its poor reconstruction, shown in Figure~\ref{fig:recons_examples},
demonstrates that the PAE model does not simultaneously
accommodate the combination of M star and galaxy.

Two other outliers exhibit problems with
the reconstruction of their continua, though
visual inspection could not confirm nor
preclude the presence of a neighboring contaminant.

\subsubsection{QSO with Foreground absorption -- Bad $z$}
Two objects are quasars misclassified by SDSS as galaxies at incorrect redshifts.
Object (53084,1440,388)
is a quasar at $z=1.838$
but has an SDSS redshift of
$z=0.05146$.
(53167,1423,287) is
at $z=1.965$
but has an SDSS redshift of
$z=0.08944$.
Absorption lines assigned to Hydrogen
as the nominal SDSS redshifts
are responsible for these misclassifications.

One of our first processing steps
is to blueshift the spectra, so it is
expected that catastrophic
redshift errors produce outlier spectra.

\subsubsection{Red continuum}
\begin{figure*}
\includegraphics[width=0.48\textwidth]{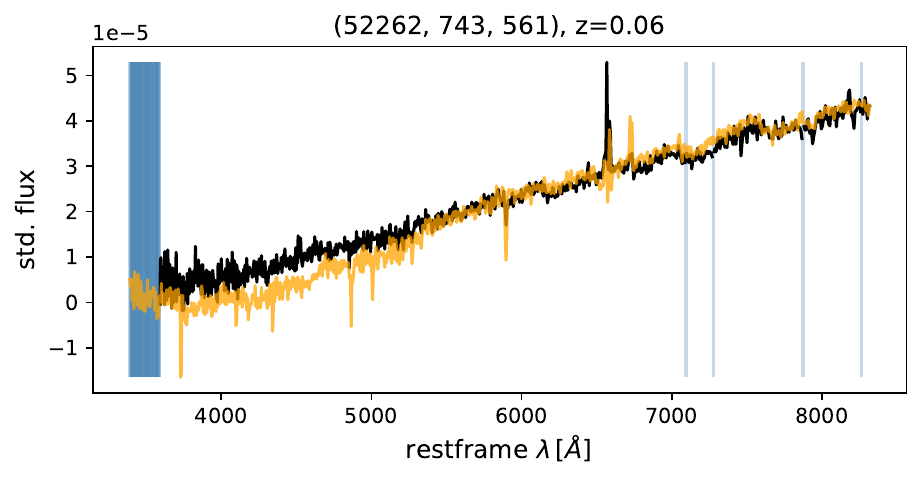}
\includegraphics[width=0.48\textwidth]{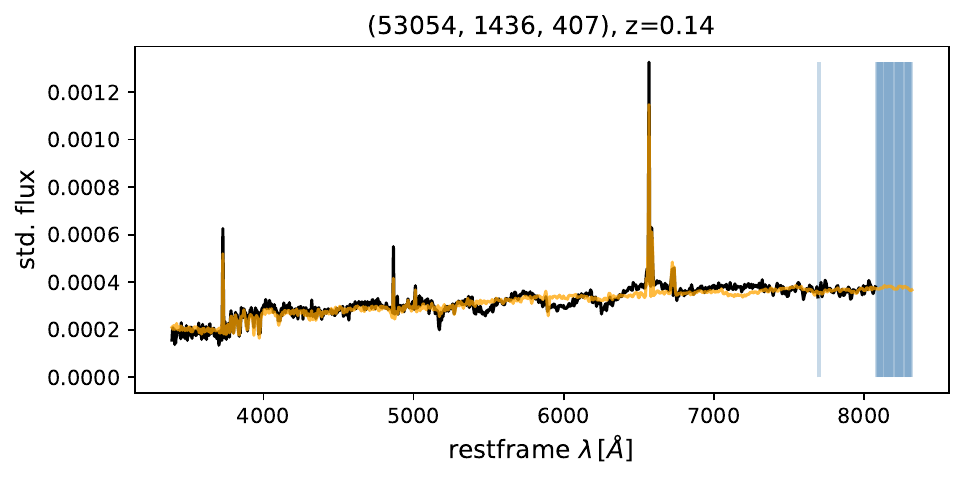}
\caption{\label{fig:recons_examples} Two objects with visually clear reconstruction error. The black curve
is the data and the orange is the reconstruction. Left: Object (52262, 743, 561)
has extreme red colors relative to the
entire SDSS sample. 
Right: Object (53054, 1436, 407) appears to include an overlap of an M star and an emission line galaxy in the same fiber. The reconstruction does not reproduce any of the M star features. Masked pixels are indicated by blue vertical lines.
}
\end{figure*}

One red galaxy, Object (52262, 743, 561) shown in Figure~\ref{fig:recons_examples},
has a continuum
that is bluer
than its PAE reconstruction. The SDSS colors of this galaxy are extremely red:
$u-g=2.8$~mag, $g-r=1.51$~mag, $g-i=2.39$~mag,
which is an extreme red outlier
\footnote{A point of comparison is the distribution in 
\url{https://www.astroml.org/examples/datasets/plot_sdss_galaxy_colors.html}.}.  
As evidenced in the reconstructed spectrum,
 the PAE can accommodate spectra
redder than the data.

\subsubsection{Supernova}

SN2001km, already identified
as highly anomalous in \S\ref{sec:sn1},
also has large reconstruction error.
In addition, Object (53770, 2376, 183) is SN~2006af
discovered by the spectroscopic search of
\citet{2013MNRAS.430.1746G}.

\subsubsection{Low-redshift, low surface brightness galaxies}
\label{recon-ston:sec}
Four spectra have reconstructions
that look nearly identical to each other but do not 
look like the data.
The spectrum and reconstruction are similar in that they
both have high-frequency `noise' fluctuations but are otherwise dissimilar in both their
broad and narrow band features.
The reconstruction error
is catastrophic, as seen by the high scores in Table B1.

\begin{figure*}
\begin{center}
\includegraphics[width=0.95\textwidth]{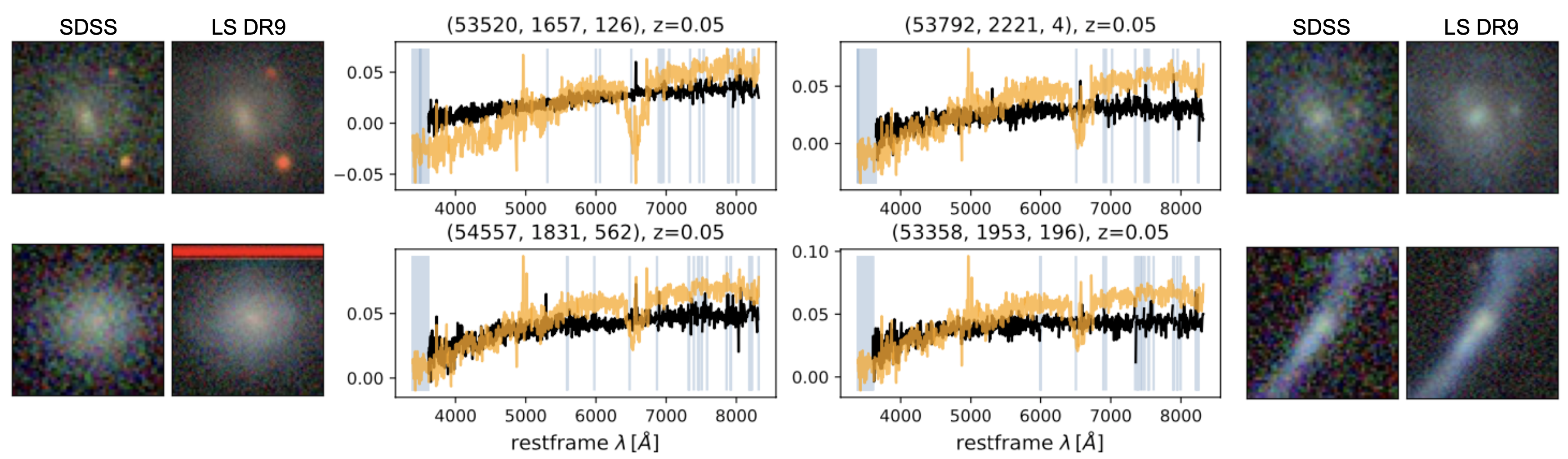}
\caption{\label{fig:4high_recon} 
Imaging and spectra of four select objects, (53520, 1656, 216), (53792, 2221, 4),
(54557, 1831, 562), and (53358, 1953, 196), that have
large reconstruction errors and similar
observed properties.
In the spectra, the black curve
is the data, the orange is the reconstruction and the blue vertical lines indicate the masked pixels.
The images are 20\arcsec $\times$ 20\arcsec cutouts from SDSS Imaging and the Legacy Survey DR9. (the SDSS fiber diameter was 2\arcsec)}
\end{center}
\end{figure*}

The imaging and spectra of these four objects share common features, as shown in Figure~\ref{fig:4high_recon}. 
Their spectra
are characterized by narrow, unresolved emission lines and their continua show a red slope with noticeable noise likely due to the low surface brightness of the galaxy.
They all share the same SDSS-determine redshift
of $z=0.05$.
The reconstructed spectra of all four objects  are similar to each other but are dissimilar to
the original data.
The images show that the galaxies are characterised by a low surface brightness with a red central region 
upon which the SDSS fiber is placed. 
Those features are unusual relative to the majority of normal low-redshift galaxies with \ha\ emission lines, which instead tend to be galaxies with star formation, blue continuum slopes and a higher surface brightness.

Recall from \S\ref{sec:contamination1} that
we found that autoencoding the output
of a previous autoencoding (Eqs.~\ref{eq:AE1}, \ref{eq:AE2}),
effectively interpolates the masked regions in the
input spectra.
Large reconstruction errors in cases
such as Object (53520, 1657, 126) do not occur
in the output of the first autoencoder.

\subsubsection{Misextraction}
Both 
(53473,   1627, 374) and
(54154, 2497, 491) have sharp negative flux lines
 and
 (52443,  986, 297)
has a deep absorption adjacent to a sharp emission
line that has been masked.
The spectrum of Object (51614, 281, 34) has asymptotically increasing red flux.
This source was observed in the 
DESI survey \citep{2016arXiv161100036D}; the spectra look similar except
that DESI does not exhibit a red runaway. 

\section{Conclusions}
\label{sec:conclusions}
We have demonstrated that a Probabilistic Autoencoder is able to learn the complex intrinsic distribution of galaxy spectra and can be used to identify rare and outlier objects
from a set of SDSS spectra.  
When examining the histogram of probabilities, we find that the model on average assigns lowest probabilities to quasars, which is expected, given their respective rareness in our training sample and large intrinsic variability. The mean probabilities of SF/AGN galaxies and quiescent galaxies are similar, but SF/AGN galaxies exhibit a longer tail to small probabilities and quiescent galaxies have the narrowest peak. 

Focusing on galaxies previously labeled as generic galaxies
by the default SDSS classifier, we find that the majority of the rare objects have significant Balmer absorption, no [OII] emission, but exhibit [NII] emission.
These are associated with the rare \EA{} galaxies with
evidence of AGN activity described by \citet{2017A&A...597A.134M}.

The PAE also identifies expected anomalies, such as blends of
multiple sources and supernovae. In addition to these expected anomalies, the PAE picked out two singular objects, one an extremely blue \EA{} galaxy and the
other a spectral mix of an early elliptical and emission-line
galaxy.

The PAE implementation used in this work includes a conditional density estimation, which can be used to incorporate data labels and additional information such as spectral classes. In reverse, this conditional model allows us to perform a probabilistic classification of unlabeled objects.
Generally, because of its probabilistic layout, the PAE can be used as part of a larger probabilistic framework, such as a Bayesian parameter inference pipeline (demonstrated in ~\cite{Stein_2022}). 

In the current form the PAE anomaly detection is not tailored toward a specific type of outlier, for example, neither the
low-probability nor outlier-reconstruction scores accentuate
the peculiarities of supernovae relative to other outliers. The method could be tailored to a specific anomaly, e.g., through calibration or suitable probabilistic conditioning. 

In principle, the ranking of rarity of the anomaly detector
is sensitive to the training data, its preprocessing, and the dimensionality of the encoded space. However, we find that the rankings for different model architectures and different random seeds are highly correlated. 

When using reconstruction errors as an outlier detection metric, we find that the results are sensitive to large point deviations away
from the learned spectral behavior. These can be caused by
misextractions and unanticipated issues with the observation itself.  Conversely,
large errors are also caused by subtle differences that persist
over large ranges of wavelength, which can be cased by a faint
background source contaminating the primary source signal.
In addition, we find that reconstruction can sometimes
fail for the noisier data in our sample.

\section*{acknowledgements}

A.G.K.\ is supported by 
the U.S.\ Department of Energy, Office of Science, Office of High Energy 
Physics, under contract No.\ DE-AC02-05CH11231. 

S.J.'s research is supported by NSF's NOIRLab, which is operated by the Association of Universities for Research in Astronomy (AURA) under a cooperative agreement with the National Science Foundation.

Funding for the SDSS and SDSS-II has been provided by the Alfred P. Sloan Foundation, the Participating Institutions, the National Science Foundation, the U.S. Department of Energy, the National Aeronautics and Space Administration, the Japanese Monbukagakusho, the Max Planck Society, and the Higher Education Funding Council for England. The SDSS Web Site is http://www.sdss.org/.

The SDSS is managed by the Astrophysical Research Consortium for the Participating Institutions. The Participating Institutions are the American Museum of Natural History, Astrophysical Institute Potsdam, University of Basel, University of Cambridge, Case Western Reserve University, University of Chicago, Drexel University, Fermilab, the Institute for Advanced Study, the Japan Participation Group, Johns Hopkins University, the Joint Institute for Nuclear Astrophysics, the Kavli Institute for Particle Astrophysics and Cosmology, the Korean Scientist Group, the Chinese Academy of Sciences (LAMOST), Los Alamos National Laboratory, the Max-Planck-Institute for Astronomy (MPIA), the Max-Planck-Institute for Astrophysics (MPA), New Mexico State University, Ohio State University, University of Pittsburgh, University of Portsmouth, Princeton University, the United States Naval Observatory, and the University of Washington.

The Legacy Surveys consist of three individual and complementary projects: the Dark Energy Camera Legacy Survey (DECaLS; Proposal ID \#2014B-0404; PIs: David Schlegel and Arjun Dey), the Beijing-Arizona Sky Survey (BASS; NOAO Prop. ID \#2015A-0801; PIs: Zhou Xu and Xiaohui Fan), and the Mayall z-band Legacy Survey (MzLS; Prop. ID \#2016A-0453; PI: Arjun Dey). DECaLS, BASS and MzLS together include data obtained, respectively, at the Blanco telescope, Cerro Tololo Inter-American Observatory, NSF’s NOIRLab; the Bok telescope, Steward Observatory, University of Arizona; and the Mayall telescope, Kitt Peak National Observatory, NOIRLab. The Legacy Surveys project is honored to be permitted to conduct astronomical research on Iolkam Du’ag (Kitt Peak), a mountain with particular significance to the Tohono O’odham Nation.

This project used data obtained with the Dark Energy Camera (DECam), which was constructed by the Dark Energy Survey (DES) collaboration. Funding for the DES Projects has been provided by the U.S. Department of Energy, the U.S. National Science Foundation, the Ministry of Science and Education of Spain, the Science and Technology Facilities Council of the United Kingdom, the Higher Education Funding Council for England, the National Center for Supercomputing Applications at the University of Illinois at Urbana-Champaign, the Kavli Institute of Cosmological Physics at the University of Chicago, Center for Cosmology and Astro-Particle Physics at the Ohio State University, the Mitchell Institute for Fundamental Physics and Astronomy at Texas A\& M University, Financiadora de Estudos e Projetos, Fundacao Carlos Chagas Filho de Amparo, Financiadora de Estudos e Projetos, Fundacao Carlos Chagas Filho de Amparo a Pesquisa do Estado do Rio de Janeiro, Conselho Nacional de Desenvolvimento Cientifico e Tecnologico and the Ministerio da Ciencia, Tecnologia e Inovacao, the Deutsche Forschungsgemeinschaft and the Collaborating Institutions in the Dark Energy Survey. The Collaborating Institutions are Argonne National Laboratory, the University of California at Santa Cruz, the University of Cambridge, Centro de Investigaciones Energeticas, Medioambientales y Tecnologicas-Madrid, the University of Chicago, University College London, the DES-Brazil Consortium, the University of Edinburgh, the Eidgenossische Technische Hochschule (ETH) Zurich, Fermi National Accelerator Laboratory, the University of Illinois at Urbana-Champaign, the Institut de Ciencies de l’Espai (IEEC/CSIC), the Institut de Fisica d’Altes Energies, Lawrence Berkeley National Laboratory, the Ludwig Maximilians Universitat Munchen and the associated Excellence Cluster Universe, the University of Michigan, NSF’s NOIRLab, the University of Nottingham, the Ohio State University, the University of Pennsylvania, the University of Portsmouth, SLAC National Accelerator Laboratory, Stanford University, the University of Sussex, and Texas A\&M University.

BASS is a key project of the Telescope Access Program (TAP), which has been funded by the National Astronomical Observatories of China, the Chinese Academy of Sciences (the Strategic Priority Research Program “The Emergence of Cosmological Structures” Grant \# XDB09000000), and the Special Fund for Astronomy from the Ministry of Finance. The BASS is also supported by the External Cooperation Program of Chinese Academy of Sciences (Grant \# 114A11KYSB20160057), and Chinese National Natural Science Foundation (Grant \# 11433005).

The Legacy Survey team makes use of data products from the Near-Earth Object Wide-field Infrared Survey Explorer (NEOWISE), which is a project of the Jet Propulsion Laboratory/California Institute of Technology. NEOWISE is funded by the National Aeronautics and Space Administration.

The Legacy Surveys imaging of the DESI footprint is supported by the Director, Office of Science, Office of High Energy Physics of the U.S. Department of Energy under Contract No. DE-AC02-05CH1123, by the National Energy Research Scientific Computing Center, a DOE Office of Science User Facility under the same contract; and by the U.S. National Science Foundation, Division of Astronomical Sciences under Contract No. AST-0950945 to NOAO.

This research has made use of the SIMBAD database, operated at CDS, Strasbourg, France. It has also made use of services and data provided by the Astro Data Lab at NSF's NOIRLab.


\section*{Data Availability}

The data underlying this article were accessed from the SDSS-BOSS DR16 release
\citep{2000AJ....120.1579Y,
2002AJ....124.1810S,
2002AJ....123.2945R,
2006AJ....131.2332G,
2020ApJS..249....3A}
available at
\url{https://skyserver.sdss.org/dr16/en/home.aspx}.
The derived data generated in this research will be shared on reasonable request to the corresponding author.

\bibliographystyle{mnras}
\bibliography{paper,ML,cosmo} 



\appendix

\section{Top Anomalies}
Table~\ref{tab:A1} presents the eight
objects for each redshift
bin with the highest anomaly score.
The peculiar
features described in \S\ref{sec:anomaly}
associated with each outlier are
given in the `explanation' column.
AGN/QSO are visually identified as an AGN or quasar.
\EA{} galaxies are characterized by Balmer absorption.
LINERS are galaxies with
a high [NII]/H$\alpha$ ratio, and generally have
no other strong emission lines.
`Neighbor` indicates the presence of multiple sources entering
the fiber.
`SN' refers to a previously identified supernova.
Low SB refers to one galaxy with a noticeably
low surface brightness and red center.
Objects with SIMBAD
LINER, AGN, or QSO classification are labeled with brackets,
e.g.\ `[AGN]'. 

\begin{table}
\scriptsize
\centering
\begin{tabular}{ccccccl}
\hline
MJD   & fiber & plate & z    & z-bin & logp   & explanation \\ \hline
52405 & 429   & 954   & 0.07 & 1     & -32.11 &   AGN.QSO, [AGN]        \\
51691 & 640   & 350   & 0.08 & 1     & -26.84 &    LINER     \\
51820 & 45    & 429   & 0.07 & 1     & -23.78 &   LINER, [AGN]\\
51955 & 247   & 472   & 0.07 & 1     & -22.17 &     SN  \\
52258 & 399   & 412   & 0.07 & 1     & -20.49 &   \EA{}, LINER, [QSO]      \\
53520 & 126   & 1657  & 0.05 & 1     & -19.38 &    Low SB        \\
53149 & 192   & 1649  & 0.07 & 1     & -19.16 &  LINER, [LINER]         \\
53084 & 388   & 1440  & 0.05 & 1     & -19.14 &  AGN/QSO,    [QSO]       \\
52232 & 83    & 762   & 0.09 & 2     & -25.72 &    \EA{}, LINER      \\
53386 & 368   & 1755  & 0.11 & 2     & -25.54 &    \EA{}, LINER     \\
54141 & 264   & 2513  & 0.11 & 2     & -25.18 &    \EA{}, LINER          \\
51882 & 11    & 435   & 0.1  & 2     & -24.36 &   \EA{}, LINER, [AGN]     \\
54561 & 586   & 1833  & 0.1  & 2     & -24.34 &   \EA{}, LINER, [AGN]   \\
53729 & 142   & 2236  & 0.09 & 2     & -22.62 &    Neighbor     \\
51984 & 374   & 498   & 0.09 & 2     & -22.52 &    \EA{}, LINER, [AGN] \\
52991 & 533   & 1270  & 0.08 & 2     & -20.4  &    \EA{}, LINER   \\
53533 & 287   & 1709  & 0.12 & 3     & -33.14 &    \EA{}, LINER, [AGN]    \\
51986 & 81    & 294   & 0.12 & 3     & -27.78 &    \EA{}, LINER    \\
53327 & 603   & 1928  & 0.13 & 3     & -26.46 &    \EA{}, LINER        \\
52964 & 91    & 1587  & 0.11 & 3     & -26.19 &    \EA{}, LINER, [LINER]      \\
53469 & 205   & 2099  & 0.13 & 3     & -25.69 &    \EA{}   \\
51873 & 487   & 443   & 0.13 & 3     & -24.35 &    \EA{}  \\
52000 & 259   & 288   & 0.12 & 3     & -24.18 &    Neighbor  \\
51930 & 256   & 285   & 0.12 & 3     & -23.84 &     \EA{}, LINER    \\
53054 & 407   & 1436  & 0.14 & 4     & -34.58 &     Neighbor      \\
53386 & 163   & 1755  & 0.14 & 4     & -29.26 &     \EA{}, LINER   \\
51930 & 115   & 285   & 0.17 & 4     & -28.88 &   LINER, [AGN]   \\
52615 & 233   & 961   & 0.16 & 4     & -28.77 &   \EA{}, LINER, [LINER]   \\
51929 & 115   & 470   & 0.16 & 4     & -28.14 &   \EA{}, LINER    \\
53089 & 114   & 1623  & 0.17 & 4     & -27.84 &    \EA{}, LINER   \\
52672 & 4     & 934   & 0.14 & 4     & -27.53 & LINER, [AGN]   \\
54589 & 77    & 2532  & 0.14 & 4     & -26.9  &     Neighbor  \\
53534 & 512   & 2112  & 0.29 & 5     & -48.2  &    \EA{}, LINER  \\
52238 & 530   & 757   & 0.29 & 5     & -40.6  &    LINER   \\
52368 & 575   & 607   & 0.24 & 5     & -39.93 &     \EA{}, LINER, [LINER] \\
52724 & 234   & 1195  & 0.23 & 5     & -39.56 &     \EA{}, LINER   \\
53875 & 327   & 1805  & 0.25 & 5     & -39.51 &   LINER    [AGN]      \\
52056 & 446   & 610   & 0.24 & 5     & -39.26 &   Neighbor  \\
53358 & 184   & 1953  & 0.26 & 5     & -38.29 &   \EA{}, LINER, [LINER]       \\
53358 & 199   & 1750  & 0.21 & 5     & -38.23 &   \EA{}, LINER, [LINER]     
\end{tabular}
\caption{\label{tab:A1} Spectra contained in the test set that were identified as anomalous because of their low probability.}
\end{table}

\section{Top Reconstruction error}
Table~\ref{tab:A2} presents the eight
objects for each redshift
bin with the highest reconstruction error.
The peculiar features described in \S\ref{sec:reconstruction}
associated with each outlier are given in the `explanation' column.
Several objects in the same exposure exhibit lines associated with `Airglow'.  
Four `Low SB' objects exhibit  
low surface brightness star forming regions and a red galaxy center leading to spectra 
with an unusual combination of a noisy red 
continuum with narrow emission lines with poor reconstructions
that do not look like the original spectra.
`QSO-Foreground' have an incorrect SDSS redshift due to foreground absorption lines.

\begin{table}
\scriptsize
\centering
\begin{tabular}{lllllll} \hline
MJD   & fiber & plate & z    & z-bin & recon error & explanation \\ \hline
51999 & 37    & 362   & 0.06 & 1     & 96.16       &  Airglow           \\
53520 & 126   & 1657  & 0.05 & 1     & 25.57       &    Low SB         \\
53792 & 4     & 2221  & 0.05 & 1     & 15.63       &    Low SB         \\
53084 & 388   & 1440  & 0.05 & 1     & 14.39       &    QSO-Foreground         \\
51955 & 247   & 472   & 0.07 & 1     & 9.93        &        SN     \\
54557 & 562   & 1831  & 0.05 & 1     & 8.94        &    Low SB         \\
53358 & 196   & 1953  & 0.05 & 1     & 8.3         &    Low SB         \\
52262 & 561   & 743   & 0.06 & 1     & 7.21        &       Red      \\
51999 & 72    & 362   & 0.1  & 2     & 20.8        &    Airglow         \\
53167 & 287   & 1423  & 0.09 & 2     & 5.46        &    QSO-Foreground       \\
53473 & 374   & 1627  & 0.09 & 2     & 4.35        &       Misextraction      \\
53729 & 142   & 2236  & 0.09 & 2     & 3.56        &       Subtle (Neighbor)      \\
53327 & 499   & 1928  & 0.08 & 2     & 3.54        &       Subtle (Neighbor)      \\
53770 & 183   & 2376  & 0.09 & 2     & 3.3         &      SN    \\
52443 & 297   & 986   & 0.09 & 2     & 3.17        &       Misextraction      \\
52441 & 60    & 790   & 0.1  & 2     & 2.93        &        Subtle (Neighbor)     \\
52000 & 259   & 288   & 0.12 & 3     & 4.44        &       Subtle (Neighbor)      \\
53494 & 379   & 1829  & 0.12 & 3     & 3.28        &       Subtle (Neighbor)      \\
51614 & 34    & 281   & 0.13 & 3     & 3.24        &       Misextraction      \\
54464 & 360   & 1876  & 0.13 & 3     & 2.97        &       Subtle (Neighbor)      \\
51999 & 273   & 362   & 0.12 & 3     & 2.8         &    Airglow         \\
51908 & 301   & 450   & 0.12 & 3     & 2.62        &    Subtle (Neighbor)         \\
53379 & 43    & 1752  & 0.12 & 3     & 2.46        &       Subtle (Neighbor)      \\
54154 & 491   & 2497  & 0.12 & 3     & 2.33        &       Misextraction      \\
51999 & 20    & 362   & 0.14 & 4     & 8.72        &    Airglow         \\
53054 & 407   & 1436  & 0.14 & 4     & 5.06        &       Neighbor      \\
51959 & 428   & 541   & 0.16 & 4     & 4.88        &       Subtle (Neighbor)      \\
51820 & 210   & 429   & 0.16 & 4     & 2.82        &       Subtle (Neighbor)      \\
53149 & 595   & 1570  & 0.15 & 4     & 2.49        &       Subtle      \\
54178 & 43    & 2492  & 0.14 & 4     & 2.43        &       Subtle (Neighbor)      \\
51993 & 166   & 334   & 0.16 & 4     & 2.41        &        Subtle     \\
52465 & 473   & 1043  & 0.14 & 4     & 2.22        &       Subtle (Neighbor)      \\
52345 & 483   & 613   & 0.18 & 5     & 3.34        &       Subtle (Neighbor)      \\
53504 & 119   & 1828  & 0.19 & 5     & 3.22        &       Misextraction      \\
54510 & 524   & 2150  & 0.18 & 5     & 3.21        &       Subtle (Neighbor)      \\
51671 & 460   & 348   & 0.21 & 5     & 3.07        &       Subtle (Neighbor)      \\
53818 & 291   & 2013  & 0.21 & 5     & 2.81        &       Subtle (Neighbor)      \\
53815 & 177   & 2427  & 0.22 & 5     & 2.65        &       Misextraction      \\
53682 & 16    & 2268  & 0.19 & 5     & 2.48        &       Subtle (Neighbor)      \\
53035 & 421   & 1735  & 0.21 & 5     & 2.44        &       Subtle (Neighbor)     
\end{tabular}
    \caption{\label{tab:A2} Spectra contained in the test set that were identified as anomalous because of their high reconstruction error}
\end{table}


\bsp	
\label{lastpage}
\end{document}